\documentclass[useAMS,usenatbib]{aastex6}
\usepackage{graphicx,amsmath,times}
\usepackage{xcolor}
\usepackage{longtable}
\usepackage{subfigure}

\begin{document}

\title{Sensitivity analysis of grain surface chemistry to binding energies of ice species}

\author{E M Penteado\altaffilmark{1}}
\author{C Walsh\altaffilmark{2, 3}}
\and
\author{H M Cuppen\altaffilmark{1}}

\email{h.cuppen@science.ru.nl}
\affil{$^1$Radboud University, Institute for Molecules and Materials, Heyendaalseweg 135, NL-6525 AJ Nijmegen, The Netherlands} 
\affil{$^2$School of Physics and Astronomy, University of Leeds, Leeds LS2 9JT, UK}
\affil{$^3$Leiden Observatory, Leiden University PO Box 9513, 2300 RA Leiden, The Netherlands}

\begin{abstract}
Advanced telescopes, such as ALMA and JWST, are likely to show that the chemical universe may be 
even more complex than currently observed, requiring astrochemical modelers to improve their models 
to account for the impact of new data. However, essential input information for gas-grain models, 
such as binding energies of molecules to the surface, have been derived experimentally only for a 
handful of species, leaving hundreds of species with highly uncertain estimates. We present in this 
paper a systematic study of the effect of uncertainties in the binding energies on an astrochemical 
two-phase model of a dark molecular cloud, using the rate equations approach. A list of recommended 
binding energy values based on a literature search of published data is presented. Thousands of 
simulations of dark cloud models were run, and in each simulation a value for the binding energy 
of hundreds of species was randomly chosen from a normal distribution. Our results show that the 
binding energy of H$_{2}$ is critical for the surface chemistry. For high binding energy, H$_2$ 
freezes out on the grain forming an H$_2$ ice. This is not physically realistic and we suggest 
a change in the rate equations. The abundance ranges found are in reasonable agreement with 
astronomical ice observations. Pearson correlation coefficients revealed that the binding energy 
of HCO, HNO, CH$_{2}$, and C correlate most strongly with the abundance of dominant ice species. 
Finally, the formation route of complex organic molecules was found to be sensitive to the branching 
ratios of H$_{2}$CO hydrogenation. 
\end{abstract}


\section{Introduction} \label{sec:intro}

The new generation of telescopes, e.g., ALMA and JWST, will provide 
us with an unprecedented wealth of molecular information. For ALMA this is at very
high spatial resolution, which depends on both the observing frequency and the maximum baseline. 
Despite the fact that ALMA observes these molecules in the
gas phase through rotational emission, many of the more complex saturated organic
molecules are thought to originate from the mantles of dust grains, since either they or their precursor
are formed through grain surface chemistry.

To interpret all this information it becomes more and more pressing to have reliable
gas-grain chemical models. Astrochemical models have developed over the decades to
understand the molecular processes in the interstellar medium (ISM). Although different
approaches have been applied, most studies use the rate equation method because of its
simplicity and its advantageous computational time. The gas phase part of these models
have been extensively reviewed and the error propagation of the input parameters in the
form of rate constants for gas phase reactions has been considered in the calculation of final abundances \citep{Wakelam:2010B}. 

Here we extend this effort to the grain surface part of gas-grain codes by looking
at the effect of uncertainties in the binding energy ($E\rm_{bind}$) of species to the grain, by applying 
the same methodology used for the gas phase reaction rates as \citet{Wakelam:2006} and \citet{Wakelam:2010B}. This
parameter determines the upper value of the temperature range at which species are
available for reactions on the grain surface. Moreover, since in most models the diffusion barrier is obtained
from the binding energy, it also, indirectly, determines the onset temperature for surface
reactions through the diffusive Langmuir-Hinshelwood mechanism.

Binding energies can be determined experimentally by the Temperature Programmed Desorption (TPD) technique, 
which will be briefly explained in the next section. But it is not straightforward to provide binding energy 
information for radical species using TPD, since these species are very short-lived under laboratory conditions. 
In some cases computational studies can provide valuable information. However, the first gas-grain chemical 
models \citep{Tielens:1982, Hasegawa:1993B, Charnley:1997C} predate TPD experiments.
The binding energies in these early works were estimated from the polarisability of the molecule 
or atom which provides an estimate of the strength of the van der Waals interaction with the bare grain surface.

In the present paper, we present an extensive literature search in which we arrive at
recommended values for the binding energies and their uncertainties. Next, thousands
of simulations are performed, where each simulation uses a set of binding energies that
is randomly picked from a Gaussian distribution considering the recommended binding
energies and their uncertainties. The species that provide the strongest effect on the
abundance of key ice components, such as H$_{2}$O, CO and CH$_{3}$OH, will be discussed and
compared against ice observations.

\section{Binding energies} \label{sec:energies}

The chemical network used in the present work \citep[][and references therein]{Drozdovskaya:2014, Drozdovskaya:2015, Garrod:2008b, Walsh:2015} 
contains 190 surface species for which
we need to provide a binding energy. As mentioned before, binding energies are typically
obtained by experiments that use the Temperature Programmed Desorption (TPD) technique. 
This well-established technique consists of two stages: first, the temperature of the substrate is kept 
constant while the species of interest is deposited; secondly, the temperature is
linearly increased until the species are desorbed from the substrate while the desorbing
species are recorded using a mass spectrometer. Next, the Polanyi-Wigner equation

\begin{equation}
 \frac{{\rm d}N}{{\rm d}t} = -k_{{m}}N^{{m}}{\rm exp} \Big[-\frac {E_{{\rm bind},i}} {k_{{\rm B}}T} \Big],
\end{equation}

\noindent where $N$ and $T$ are respectively the total number of absorbed species and the surface temperature 
both at a certain time $t$, is fitted to the measured desorption spectra of a species $i$ to obtain $E\rm_{bind,i}$, a 
prefactor, and in some cases also the desorption order $m$. Since the prefactor and binding energy are highly correlated,
several experiments need to be performed with different amounts of predeposited species.
The final results depend on the nature of the substrate from which the species desorbs.
As an example we would like to mention the reported experimental values of $E\rm_{bind}$ for
water, which may range from 4800 K, as reported by \cite{Dulieu:2013} from studies of formation of
water on amorphous silicate surfaces, to 5930$\pm$240 K, as reported by \cite{Collings:2015}, based on experiments of
desorption from amorphous silica. Whether experiments are performed in the
monolayer or multilayer regime and whether the deposited ice is pure, mixed, or layered
also has an effect. In the multilayer regime, desorption occurs from the species itself and
the effect of the underlying substrate becomes negligible as pointed out by \cite{Green:2009}. Most experiments 
concentrate their attention on a few important species, such as H$_{2}$O, CO, CO$_{2}$, and CH$_{3}$OH. Data 
on more complex species is far more sparse. Our collected data is presented in Table~\ref{experiment}. 
Here we take the binding energies from experiments with an amorphous water ice substrate 
where possible, as an attempt to use homogeneous data, since the differences between different substrates 
can be large, although smaller in the multilayer regime \citep{Green:2009}. The quoted uncertainties 
in the table are a combination of experimental 
errors, fitting errors and intrinsic variety of the binding energies due to the inhomogeneity 
of the substrate. The latter is especially important in the case of amorphous substrates. 

In some cases, computational studies can help to derive binding energies. For instance, 
\cite{Al-Halabi:2007} simulated adsorption of H atoms to amorphous solid water using classical
trajectory calculations. The off-lattice kinetic Monte Carlo approach was used by \cite{Karssemeijer:2014III}
to estimate binding energies of CO and CO$_{2}$. For the majority of species, however, there
are still no laboratory data available nor computational estimates.
{\small
\begin{table*}
\caption{Experimentally determined $E_\text{bind}$ values and specifics of each experiment.}
 \begin{center}
  \begin{tabular}{l r@{ $\pm$ }r c r l c l r}
  \hline
  \hline
  Species & \multicolumn{2}{c}{$E_\text{bind}$ (K)} & Order & Prefactor$^a$ & Substrate & Technique& Ref. \\
  \hline
  H$_{2}$        &  480  & 10 &   1 & $2\times 10^{12}$ & Submonolayer on amorphous silicate& TPD& 1\\[0.1cm]
  O$_{2}$        & \multicolumn{2}{c}{914--1161} & 1  & $10^{12}$ & Submonolayer on non-porous ASW& TPD& 2\\[0.1cm]
                 & 1100 &    & 1  & $5.4\times 10^{14}$  & Submonolayer on non-porous ASW& TPD & 3\\[0.1cm]
  N$_2$          & 1200 &    & 1  & $4.1\times 10^{15}$  & Submonolayer on non-porous ASW& TPD & 3\\[0.1cm]
  CO             &  \multicolumn{2}{c}{863--1307}     & 1  & $10^{12}$  & Submonolayer on non-porous ASW& TPD & 2\\[0.1cm]
                 &  1420 &    & 1  & $3.5\times 10^{16}$  & Submonolayer on non-porous ASW& TPD & 3\\[0.1cm]
  CH$_4$         &  1370 &    & 1  & $9.8\times 10^{14}$  & Submonolayer on non-porous ASW& TPD & 3\\[0.1cm]
  O              &  1764 & 232 & 1 & $10^{12}$ & Submonolayer on amorphous silicate& TPD& 3a\\[0.1cm]
  O$_3$          & 2100  &         & 1 & $10^{12}$ & Submonolayer on amorphous silicate & TPD &  5\\[0.1cm]
  OH             &  \multicolumn{2}{c}{1656--4760} & 1 & $10^{12}$ & Formed on amorphous silicate& TPD& 6\\[0.1cm]
  CO$_{2}$       &  \multicolumn{2}{c}{2236--2346}  & 1  & $10^{12}$ & Submonolayer on non-porous ASW& TPD& 2\\[0.1cm]
  C$_2$H$_6$     &  2490 && 1  & $9.5\times 10^{15}$  & Submonolayer on non-porous ASW& TPD & 3\\[0.1cm]
  H$_{2}$CO      & 3260  & 60 & 0               & $10^{28}$ & Submonolayer on non-porous ASW& TPD&  7\\[0.1cm]
  H$_{2}$O       & 4815  & 15      &   1 & $2\times 10^{12}$ & Annealed multilayer on CsI& IR spec & 8\\
                 & 5070  & 50      &   1 & $2\times 10^{12}$ & Unannealed multilayer on CsI& IR spec & 8\\
                 & 5600  &       & 0     & $1\times 10^{30}$ & Amorphous multilayer on gold & TPD & 9\\
                 & 4799  &  96     & 0.26 $\pm$ 0.02 & $1\times 10^{27\pm 1}$ & Crystalline multilayer on HOPG & TPD& 10\\
                 & 5930  & 240     & 0               & $10^{28}$ & Multilayer on amorphous silica & TPD &11\\
                 & 4800  &         & 1 & $10^{12}$ & Submonolayer on amorphous silicate & TPD &  5\\
  H$_{2}$O$_2$   & 6000  &         & 1 & $10^{12}$ & Submonolayer on amorphous silicate & TPD &  5\\[0.1cm]
  
  \hline
  \end{tabular} \\[0.1cm]
 \end{center}
  \footnotesize{References: 
  $^a$ In units s$^{-1}$ for first-order desorption and molec cm$^{-2}$ s$^{-1}$ for zeroth-order desorption. Fractional orders are converted to nearest integer value desorption.}
1~--~\cite{Acharyya:2014} 
2~--~\cite{Noble:2012a} 
3~--~\cite{He:2014} 
3~--~\cite{Smith:2016} 
5~--~\cite{Dulieu:2013} 
6~--~\cite{He:2014A}
7~--~\cite{Noble:2012b} 
8~--~\cite{Sandford:1988a} 
9~--~\cite{Fraser:2001} 
10~--~\cite{Brown:2007} 
11~--~\cite{Collings:2015} \\
 \label{experiment}
\end{table*}
}

{\small
\begin{table*}
\caption{Estimated $T\rm{_{des}}$, and $E\rm_{bind}$ from TPD experiments.}
 \begin{center}
  \begin{tabular}{l r r@{ $\pm$ }r }
  \hline
  \hline
  Species & \multicolumn{1}{c}{$T\rm{_{des}}$ (K)} & \multicolumn{2}{c}{$E\rm_{bind}$ (K)}  \\
  \hline
  \hline
O$_{2}$        & 29  &  865 & 55$^a$ \\
               & 45  & 1342 & 65$^b$ \\[0.1cm]
N$_{2}$        & 30  &  895 & 55$^a$ \\
               & 46  & 1305 & 70$^b$ \\[0.1cm]
C$_{2}$H$_{2}$ & 70  & 2090 & 85 \\[0.1cm]
H$_{2}$S       & 77  & 2296 & 90 \\[0.1cm] 
CO$_{2}$       & 78  & 2325 & 95 \\[0.1cm]
OCS            & 78  & 2325 & 95 \\[0.1cm]
NH$_{3}$       & 91  & 2715 & 105 \\[0.1cm]
CS$_{2}$       & 95  & 2832 & 105 \\[0.1cm]
SO$_{2}$       & 101 & 3010 & 110 \\[0.1cm]
CH$_{3}$CN     & 127 & 3790 & 130 \\[0.1cm]
CH$_{3}$OH     & 128 & 3820 & 135 \\[0.1cm]
HCOOH          & 152 & 4532 & 150 \\[0.1cm]
\hline
  \end{tabular} \\[0.1cm]
   $^a$\footnotesize{First desorption peak.}  \\
   $^b$\footnotesize{Second desorption peak.} \\
 \end{center}
 \label{estimation}
\end{table*}
}

In a few experimental studies, TPD spectra are presented in figures, but no binding
energy data is provided by the authors. An example is the paper by \cite{Collings:2004}. They present an
extensive experimental study of a collection of astrophysically relevant molecular species
using the TPD technique. Three kinds of experiments were performed, differentiated
according to the substrate used: 1) deposition of each species on a pure gold substrate; 2)
deposition on a H$_{2}$O substrate; and 3) co-deposition of each species with water forming
a mixture as a substrate. \cite{Collings:2004} classified the studied species into three categories based
on their desorption behavior. One of these categories is the defined CO-like species, composed of N$_2$, O$_2$, CH$_4$.
These very volatile species desorb similarly to CO, presenting two desorption peaks. This is especially evident for N$_{2}$ and
O$_{2}$. Molecules in the second category are H$_{2}$O-like, showing a desorption behavior similar to H$_{2}$O. Included in this category are
NH$_{3}$, CH$_{3}$OH, and HCOOH. Different from the most volatile species, these H$_2$O-like species are unable to diffuse easily, and it 
is likely that these species bind strongly to the H$_2$O molecules. Finally, we have the intermediate species category. In this 
category we can find H$_{2}$S, OCS, CO$_{2}$, C$_{2}$H$_{2}$, SO$_{2}$, CS$_{2}$, and CH$_{3}$CN. When desorbing from a water ice, these 
species present a volcano desorption and co-desorption peak, indicating that they are trapped in the water ice. Here, we have estimated
the binding energies based on the results presented in \cite{Collings:2004} for the deposition of each 
of those species on a H$_{2}$O substrate. Our estimation is very simple and was done as follows:

\begin{equation}
 E_{\rm{bind,X}} = \frac{T_\text{des,X}}{T_{\rm{des,H_{2}O}}} \times
E_{\rm{bind,H_{2}O}}
\end{equation}

\noindent where $T\rm{_{des,X}}$ is the desorption temperature of species X deposited on a H$_{2}$O film, $T\rm_{des,H_{2}O}$ is the 
desorption temperature of H$_{2}$O, and $E\rm_{bind,H_{2}O}$ is the binding energy of H$_{2}$O. The binding energy of water has been 
determined for a range of different substrates and preparation methods for the water ice. Here we chose 
$E\rm_{bind,H_{2}O}$ = 4800 K, which is the result of a compilation of the amorphous water results 
\citep{Sandford:1988a, Fraser:2001, Brown:2007, Collings:2015} taking into account the prefactor used in the
gas-grain model. It agrees closely with the water binding energy to amorphous silicate \citep{Dulieu:2013}. 
The values of $T\rm_{des,H_{2}O}$ and $T\rm{_{des,X}}$ were 
obtained by visual inspection of Figure 2 of \cite{Collings:2004}, from where $T\rm_{des,H_{2}O}$ was 
estimated as 161 K, and $T\rm{_{des,X}}$ varies according to the species.  Since our approach results 
in an inherent error in the estimated binding energy, we applied an uncertainty of 3.5~\% to values 
in Table~\ref{estimation}. As can be seen from Table~\ref{experiment}, experimental binding energies 
for stable species typically have an error around 1~\%. 

The values we estimated for $E\rm_{bind,CO}$, $E\rm_{bind,CO_{2}}$ and $E\rm_{bind,O_{2}}$
are in close agreement with those found by \cite{Noble:2012a} for different substrates. Our 
estimations for $E\rm_{bind,N_{2}}$, based on 
\cite{Collings:2004} using the first peak, are similar to those found in other 
studies \citep{Fuchs:2006, Oberg:2005, Collings:2015}. The value found by \cite{Smith:2016} is 
much larger, but this can be explained by the much larger pre-exponential factor that they obtained 
in their fit. If this is accounted for by assuming the same desorption rate at the peak temperature 
($\approx 25$~K), the binding energy corresponds to 990~K which is in closer agreement with our 
estimation from \cite{Collings:2004}. The reported binding energies for carbon monoxide, on the other hand, 
show a large deviation. \cite{Oberg:2005}, for instance, found $E\rm_{bind,CO}$ = 855 K based on 
experiments of multilayers of pure CO ice, while 
computational binding energies of \cite{Karssemeijer:2014III} of CO on various different water 
substrate 
ranged between 1555 K and 1700 K. Other experimental studies 
\citep{Sandford:1988b, Acharyya:2007, Collings:2003b, Collings:2015, Fuchs:2006, Noble:2012a, Smith:2016} 
obtained values within these extremes. 
Our recommended value is based on \cite{Noble:2012a}, which is representative 
of CO on water ice surface in the submonolayer regime. Additionally, we use a relatively high value of 
250~K for the uncertainty to  account for the large distribution of binding sites found in this study.

\begin{figure}
\begin{center}
  \centering
  \includegraphics[width=0.5\textwidth]{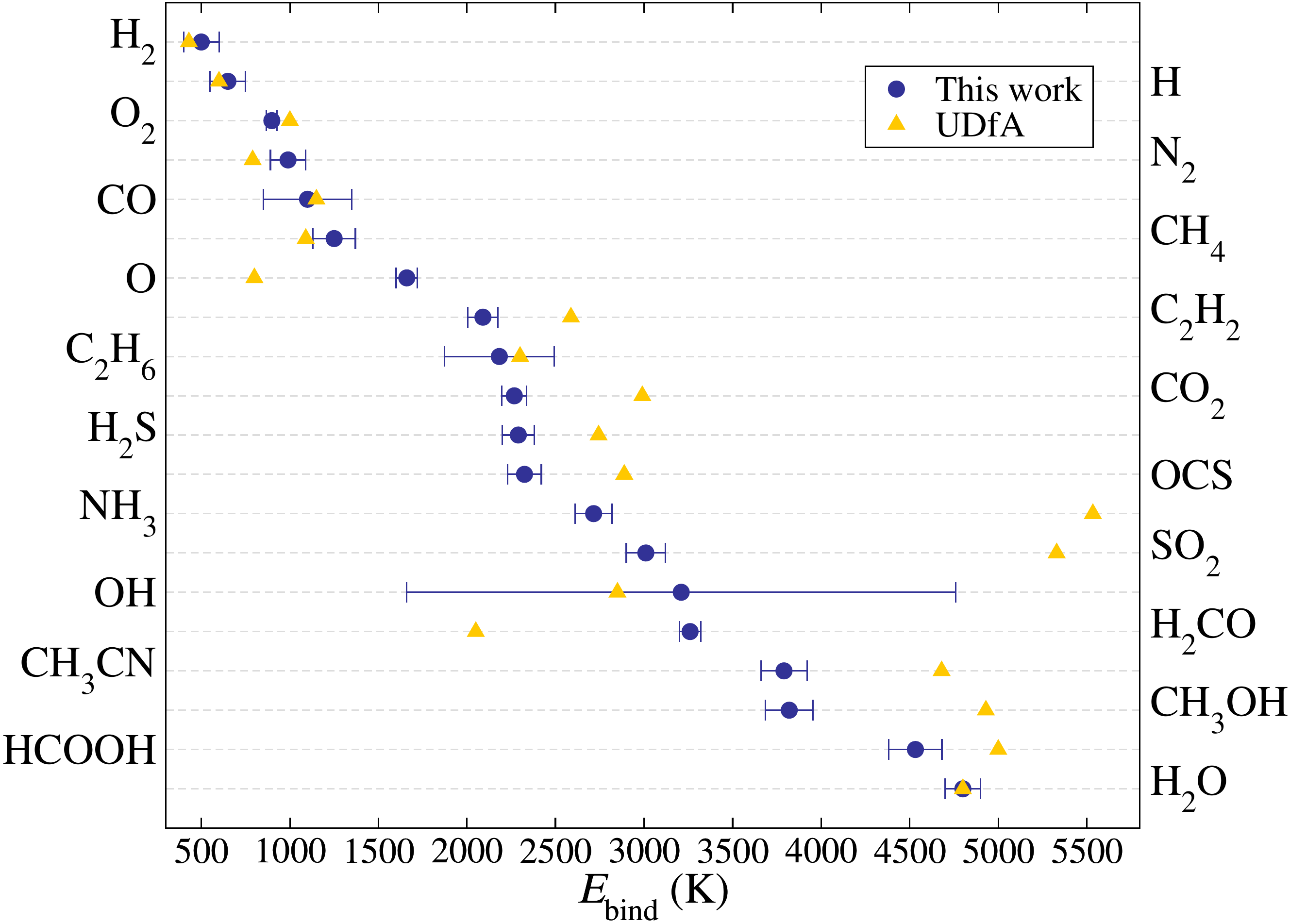}
   \caption{Comparison of binding energies recommended by the present work (blue circles) based on an extensive literature 
   review and those binding energies currently recommended by UDfA (yellow triangles). Most of species shown here are those for which experimental 
   values have been derived. The dashed gray line is a guide to the eye to identify the correspondent species.}
     \label{comparison}
\end{center}
\end{figure}

The binding energies and the correspondent uncertainties used here are summarized in 
Table~\ref{energies}. Experimental binding energy values are used for the following species: 
C$_{2}$H$_{2}$, CH$_{3}$CN, CH$_{3}$OH, CH$_{4}$, CO, CO$_{2}$, H$_{2}$, H$_{2}$CO, H$_{2}$O, 
H$_{2}$S, HCOOH, N$_{2}$, NH$_{3}$, O, O$_{2}$, OCS, OH, and SO$_{2}$. \cite{He:2014A} could only 
constrain the binding energy of OH to fall between 1656 K and 4760 K. 
We used here the average with an uncertainty to cover this range. For H$_{2}$, experiments 
deliver a large spread in the results. We chose to use here the intermediate value 
$E\rm_{bind,H_{2}}$~=~500~K,
which is very similar to the value found by \cite{Acharyya:2014} using TPD experiments. The value for 
CH$_4$ is based on \cite{Smith:2016} who found a value of 1370~K with a high prefactor. Applying a 
prefactor of $10^{12}$~s$^{-1}$ which is more representative of the prefactor typically used in 
gas-grain codes, an assuming the same desorption rate at the peak intensity (35~K), a binding 
energy of 1130~K is obtained. We use an intermediate value of 1250~K with an uncertainty of 120~K. 
The same approach is applied for C$_2$H$_4$.
Table~\ref{energies} also shows the binding energies 
recommended by the UMIST\footnote{http://www.udfa.net} Database for Astrochemistry UDfA 
\citep{McElroy:2013}, which is commonly used. It is clear 
that for most of the species the values of binding energies recommended by the present work 
and by the UDfA present large 
differences. These differences are better visualized in Figure~\ref{comparison}, where the 
values of binding energies recommended by this work and 
by the UDfA database are plotted in a common diagram for a selection of species. For some 
species these differences reflect a different choice in substrate. 
For other cases the UDfA list simply predates the experimental work on which our value is 
based.

For all other species, we adopt the initial list from \cite{Hasegawa:1993A},
which in turn is based on previous work \citep{Allen:1977, Tielens:1987}, \cite{Aikawa:1996}, 
and from \cite{Garrod:2006a}. For these cases an uncertainty of 500~K has been used when the  value 
is higher than 1000~K, otherwise the uncertainty is set to half of the binding energy.

  \begin{longtable}{ll@{ $\pm$ }lr || ll@{ $\pm$ }lr || ll@{ $\pm$ }lr}
  \caption{Molecular binding energies. Species for which experimental values have been derived are highlighted.}\\
\hline \hline
              & \multicolumn{3}{c}{$E_\text{bind}$ (K)} & & \multicolumn{3}{c}{$E_\text{bind}$ (K)} & & \multicolumn{3}{c}{$E_\text{bind}$ (K)}  \\
    {Species} & \multicolumn{2}{c}{this work} & UMIST &
    {Species} & \multicolumn{2}{c}{this work} & UMIST &
    {Species} & \multicolumn{2}{c}{this work} & UMIST \\
    \hline \hline 
\endfirsthead
\multicolumn{9}{c}{\normalsize {\tablename} \thetable\~--~\textit{Continued from previous page} --}\\
  \hline\hline 
              & \multicolumn{3}{c}{$E_\text{bind}$ (K)} & & \multicolumn{3}{c}{$E_\text{bind}$ (K)} & & \multicolumn{3}{c}{$E_\text{bind}$ (K)}  \\
    {Species} & \multicolumn{2}{c}{this work} & UMIST &
    {Species} & \multicolumn{2}{c}{this work} & UMIST &
    {Species} & \multicolumn{2}{c}{this work} & UMIST \\
    \hline\hline 
\endhead

 \textcolor{gray}{C                } & \textcolor{gray}{715}  & \textcolor{gray}{360$^f$              } & \textcolor{gray}{800}  & \textcolor{gray}{CH$_{2}$OHCHO     } & \textcolor{gray}{6680} & \textcolor{gray}{500                     } & \textcolor{gray}{    } & \textcolor{gray}{HNO             } & \textcolor{gray}{1510} & \textcolor{gray}{500$^d$              } & \textcolor{gray}{2050} \\  
 \textcolor{gray}{C$_{2}$          } & \textcolor{gray}{1085} & \textcolor{gray}{500$^f$              } & \textcolor{gray}{1600} & \textcolor{gray}{CH$_{2}$OHCO      } & \textcolor{gray}{6230} & \textcolor{gray}{500                     } & \textcolor{gray}{    } & \textcolor{gray}{HNOH            } & \textcolor{gray}{5230} & \textcolor{gray}{500                  } & \textcolor{gray}{    } \\ 
 \textcolor{gray}{C$_{2}$H         } & \textcolor{gray}{1330} & \textcolor{gray}{500$^f$              } & \textcolor{gray}{2137} & \textcolor{gray}{CH$_{2}$PH        } & \textcolor{gray}{1200} & \textcolor{gray}{500                     } & \textcolor{gray}{    } & \textcolor{gray}{HNSi            } & \textcolor{gray}{1100} & \textcolor{gray}{500                  } & \textcolor{gray}{1100} \\ 
 C$_{2}$H$_{2}$                      & 2090                   &  \hspace{1 mm} 85$^a$                   & 2587                   & \textcolor{gray}{CH$_{3}$          } & \textcolor{gray}{1040} & \textcolor{gray}{500$^f$                 } & \textcolor{gray}{1175} & \textcolor{gray}{HOCN            } & \textcolor{gray}{2850} & \textcolor{gray}{500                  } & \textcolor{gray}{    } \\ 
 \textcolor{gray}{C$_{2}$H$_{3}$   } & \textcolor{gray}{1760} & \textcolor{gray}{500$^d$              } & \textcolor{gray}{3037} & \textcolor{gray}{CH$_{3}$C$_{3}$N  } & \textcolor{gray}{3880} & \textcolor{gray}{500$^d$                 } & \textcolor{gray}{6480} & \textcolor{gray}{HONC            } & \textcolor{gray}{2850} & \textcolor{gray}{500                  } & \textcolor{gray}{    } \\ 
 \textcolor{gray}{C$_{2}$H$_{4}$   } & \textcolor{gray}{2010} & \textcolor{gray}{500$^d$              } & \textcolor{gray}{3487} & \textcolor{gray}{CH$_{3}$C$_{4}$H  } & \textcolor{gray}{3830} & \textcolor{gray}{500$^d$                 } & \textcolor{gray}{5887} & \textcolor{gray}{HPO             } & \textcolor{gray}{1200} & \textcolor{gray}{500                  } & \textcolor{gray}{    } \\ 
 \textcolor{gray}{C$_{2}$H$_{4}$CN } & \textcolor{gray}{5930} & \textcolor{gray}{500                  } & \textcolor{gray}{    } & \textcolor{gray}{CH$_{3}$C$_{5}$N  } & \textcolor{gray}{5080} & \textcolor{gray}{500$^d$                 } & \textcolor{gray}{7880} & \textcolor{gray}{HS              } & \textcolor{gray}{1350} & \textcolor{gray}{500$^f$              } & \textcolor{gray}{1500} \\ 
 \textcolor{gray}{C$_{2}$H$_{5}$   } & \textcolor{gray}{2110} & \textcolor{gray}{500$^d$              } & \textcolor{gray}{3937} & \textcolor{gray}{CH$_{3}$C$_{6}$H  } & \textcolor{gray}{5030} & \textcolor{gray}{500$^d$                 } & \textcolor{gray}{7487} & \textcolor{gray}{HS$_{2}$        } & \textcolor{gray}{2300} & \textcolor{gray}{500$^d$              } & \textcolor{gray}{2650} \\ 
 \textcolor{gray}{C$_{2}$H$_{5}$CN } & \textcolor{gray}{6380} & \textcolor{gray}{500                  } & \textcolor{gray}{    } & \textcolor{gray}{CH$_{3}$C$_{7}$N  } & \textcolor{gray}{6290} & \textcolor{gray}{500$^d$                 } & \textcolor{gray}{9480} & \textcolor{gray}{HCl             } & \textcolor{gray}{ 900} & \textcolor{gray}{450                  } & \textcolor{gray}{ 900} \\ 
 \textcolor{gray}{C$_{2}$H$_{5}$OH } & \textcolor{gray}{3470} & \textcolor{gray}{500$^d$              } & \textcolor{gray}{5200} & \textcolor{gray}{CH$_{3}$CCH       } & \textcolor{gray}{4290} & \textcolor{gray}{500                     } & \textcolor{gray}{2470} & \textcolor{gray}{F               } & \textcolor{gray}{ 450} & \textcolor{gray}{225                  } & \textcolor{gray}{    } \\ 
 C$_{2}$H$_{6}$    & 2183 & 310$^b$               & {2300} & \textcolor{gray}{CH$_{3}$CHCH$_{2}$} & \textcolor{gray}{5190} & \textcolor{gray}{500                     } & \textcolor{gray}{    } & \textcolor{gray}{Fe              } & \textcolor{gray}{3750} & \textcolor{gray}{500$^f$              } & \textcolor{gray}{4200} \\ 
 \textcolor{gray}{C$_{2}$N         } & \textcolor{gray}{2010} & \textcolor{gray}{500$^d$              } & \textcolor{gray}{2400} & \textcolor{gray}{CH$_{3}$CHO       } & \textcolor{gray}{2870} & \textcolor{gray}{500$^d$                 } & \textcolor{gray}{3800} & \textcolor{gray}{Mg              } & \textcolor{gray}{4750} & \textcolor{gray}{500$^f$              } & \textcolor{gray}{5300} \\ 
 \textcolor{gray}{C$_{2}$O         } & \textcolor{gray}{2010} & \textcolor{gray}{500$^d$              } & \textcolor{gray}{1950} & CH$_{3}$CN                           & 3790                   & 130$^a$                                    & 4680                   & \textcolor{gray}{N               } & \textcolor{gray}{ 715} & \textcolor{gray}{358$^f$              } & \textcolor{gray}{ 800} \\ 
 \textcolor{gray}{C$_{2}$S         } & \textcolor{gray}{2500} & \textcolor{gray}{500$^d$              } & \textcolor{gray}{    } & \textcolor{gray}{CH$_{3}$CO        } & \textcolor{gray}{2320} & \textcolor{gray}{500                     } & \textcolor{gray}{    } & N$_{2}$                            &  990                   &   100$^b$                   &  790                   \\ 
 \textcolor{gray}{C$_{3}$          } & \textcolor{gray}{2010} & \textcolor{gray}{500$^d$              } & \textcolor{gray}{2400} & \textcolor{gray}{CH$_{3}$COCH$_{3}$} & \textcolor{gray}{3300} & \textcolor{gray}{500                     } & \textcolor{gray}{    } & \textcolor{gray}{N$_{2}$O        } & \textcolor{gray}{2400} & \textcolor{gray}{500                  } & \textcolor{gray}{2400} \\ 
 \textcolor{gray}{C$_{3}$H         } & \textcolor{gray}{2270} & \textcolor{gray}{500$^d$              } & \textcolor{gray}{2937} & \textcolor{gray}{CH$_{3}$COOH      } & \textcolor{gray}{6300} & \textcolor{gray}{500                     } & \textcolor{gray}{    } & \textcolor{gray}{NCCN            } & \textcolor{gray}{1300} & \textcolor{gray}{500                  } & \textcolor{gray}{1300} \\ 
 \textcolor{gray}{C$_{3}$H$_{2}$   } & \textcolor{gray}{2110} & \textcolor{gray}{500$^d$              } & \textcolor{gray}{3387} & \textcolor{gray}{CH$_{3}$NH        } & \textcolor{gray}{1760} & \textcolor{gray}{500$^d$                 } & \textcolor{gray}{3553} & \textcolor{gray}{NH              } & \textcolor{gray}{ 542} & \textcolor{gray}{270$^f$              } & \textcolor{gray}{2378} \\ 
 \textcolor{gray}{C$_{3}$N         } & \textcolor{gray}{2720} & \textcolor{gray}{500$^d$              } & \textcolor{gray}{3200} & \textcolor{gray}{CH$_{3}$NH$_{2}$  } & \textcolor{gray}{5130} & \textcolor{gray}{500                     } & \textcolor{gray}{    } & \textcolor{gray}{NH$_{2}$        } & \textcolor{gray}{ 770} & \textcolor{gray}{385$^f$              } & \textcolor{gray}{3956} \\ 
 \textcolor{gray}{C$_{3}$O         } & \textcolor{gray}{2520} & \textcolor{gray}{500$^d$              } & \textcolor{gray}{2750} & \textcolor{gray}{CH$_{3}$O         } & \textcolor{gray}{2655} & \textcolor{gray}{500                     } & \textcolor{gray}{5080} & \textcolor{gray}{NH$_{2}$CHO     } & \textcolor{gray}{5560} & \textcolor{gray}{500                  } & \textcolor{gray}{5556} \\ 
 \textcolor{gray}{C$_{3}$P         } & \textcolor{gray}{1650} & \textcolor{gray}{500                  } & \textcolor{gray}{    } & \textcolor{gray}{CH$_{3}$OCH$_{3}$ } & \textcolor{gray}{2820} & \textcolor{gray}{500$^d$                 } & \textcolor{gray}{3300} & \textcolor{gray}{NH$_{2}$CN      } & \textcolor{gray}{1200} & \textcolor{gray}{500                  } & \textcolor{gray}{1200} \\ 
 \textcolor{gray}{C$_{3}$S         } & \textcolor{gray}{3000} & \textcolor{gray}{500$^d$              } & \textcolor{gray}{3500} & CH$_{3}$OH                           & 3820                   & 135$^a$                                    & 4930                   & \textcolor{gray}{NH$_{2}$CO      } & \textcolor{gray}{5110} & \textcolor{gray}{500                  } & \textcolor{gray}{    } \\ 
 \textcolor{gray}{C$_{4}$          } & \textcolor{gray}{2420} & \textcolor{gray}{500$^d$              } & \textcolor{gray}{3200} & CH$_{4}$                             & 1250                   &  120$^b$                      & 1090                   & \textcolor{gray}{NH$_{2}$OH      } & \textcolor{gray}{2770} & \textcolor{gray}{500$^d$              } & \textcolor{gray}{6806} \\ 
 \textcolor{gray}{C$_{4}$H         } & \textcolor{gray}{2670} & \textcolor{gray}{500$^d$              } & \textcolor{gray}{3737} & \textcolor{gray}{CN                } & \textcolor{gray}{1355} & \textcolor{gray}{500$^f$                 } & \textcolor{gray}{1600} & NH$_{3}$                           & 2715                   & 105$^a$                                 & 5534                   \\ 
 \textcolor{gray}{C$_{4}$H$_{2}$   } & \textcolor{gray}{2920} & \textcolor{gray}{500$^d$              } & \textcolor{gray}{4187} & \textcolor{gray}{CNO               } & \textcolor{gray}{2400} & \textcolor{gray}{500                     } & \textcolor{gray}{    } & \textcolor{gray}{NO              } & \textcolor{gray}{1085} & \textcolor{gray}{500$^f$              } & \textcolor{gray}{1600} \\ 
 \textcolor{gray}{C$_{4}$H$_{3}$   } & \textcolor{gray}{2970} & \textcolor{gray}{500$^d$              } & \textcolor{gray}{4637} & CO                                   & 1100                   & 250$^j$                                    & 1150                   & \textcolor{gray}{NO$_{2}$        } & \textcolor{gray}{2400} & \textcolor{gray}{500                  } & \textcolor{gray}{2400} \\ 
 \textcolor{gray}{C$_{4}$H$_{6}$   } & \textcolor{gray}{5990} & \textcolor{gray}{500                  } & \textcolor{gray}{5987} & CO$_{2}$                             & 2267                   & \hspace{1 mm} 70$^j$                       & 2990                   & \textcolor{gray}{NS              } & \textcolor{gray}{1800} & \textcolor{gray}{500$^f$              } & \textcolor{gray}{1900} \\ 
 \textcolor{gray}{C$_{4}$N         } & \textcolor{gray}{3220} & \textcolor{gray}{500$^d$              } & \textcolor{gray}{4000} & \textcolor{gray}{COOCH$_{3}$       } & \textcolor{gray}{3650} & \textcolor{gray}{500                     } & \textcolor{gray}{    } & \textcolor{gray}{Na              } &\textcolor{gray}{10600} & \textcolor{gray}{500$^f$              } &\textcolor{gray}{11800} \\ 
 \textcolor{gray}{C$_{4}$P         } & \textcolor{gray}{1950} & \textcolor{gray}{500                  } & \textcolor{gray}{    } & \textcolor{gray}{COOH              } & \textcolor{gray}{5120} & \textcolor{gray}{500                     } & \textcolor{gray}{    } & O                                  & 1660                   & \hspace{1 mm} 60$^i$                    &                    800 \\ 
 \textcolor{gray}{C$_{4}$S         } & \textcolor{gray}{3500} & \textcolor{gray}{500$^d$              } & \textcolor{gray}{4300} & \textcolor{gray}{CP                } & \textcolor{gray}{1050} & \textcolor{gray}{500                     } & \textcolor{gray}{    } & O$_{2}$                            &  898                   & \hspace{1 mm} 30$^j$                    &                   1000 \\ 
 \textcolor{gray}{C$_{5}$          } & \textcolor{gray}{3220} & \textcolor{gray}{500$^d$              } & \textcolor{gray}{4000} & \textcolor{gray}{CS                } & \textcolor{gray}{1800} & \textcolor{gray}{500$^f$                 } & \textcolor{gray}{1900} & \textcolor{gray}{O$_{2}$H        } & \textcolor{gray}{1510} & \textcolor{gray}{500$^d$              } & \textcolor{gray}{3650} \\ 
 \textcolor{gray}{C$_{5}$H         } & \textcolor{gray}{3470} & \textcolor{gray}{500$^d$              } & \textcolor{gray}{4537} & \textcolor{gray}{Cl                } & \textcolor{gray}{ 850} & \textcolor{gray}{425                     } & \textcolor{gray}{ 850} & O$_{3}$                            & 2100                   & 100$^c$                                 & 1800                   \\ 
 \textcolor{gray}{C$_{5}$H$_{2}$   } & \textcolor{gray}{3730} & \textcolor{gray}{500$^d$              } & \textcolor{gray}{4987} & \textcolor{gray}{ClO               } & \textcolor{gray}{1250} & \textcolor{gray}{500                     } & \textcolor{gray}{1250} & \textcolor{gray}{OCN             } & \textcolor{gray}{1805} & \textcolor{gray}{500$^f$              } & \textcolor{gray}{2400} \\ 
 \textcolor{gray}{C$_{5}$N         } & \textcolor{gray}{3930} & \textcolor{gray}{500$^d$              } & \textcolor{gray}{4800} & \textcolor{gray}{H                 } & \textcolor{gray}{ 650} & \textcolor{gray}{100$^k$                 } & \textcolor{gray}{ 600} & OCS                                & 2325                   & \hspace{1 mm} 95$^a$                    & 2888                   \\ 
 \textcolor{gray}{C$_{6}$          } & \textcolor{gray}{3620} & \textcolor{gray}{500$^d$              } & \textcolor{gray}{4800} & H$_{2}$                              & 500                    & 100$^e$                                    & 430                    & OH                                 & 3210                   & 1550$^i$                                & 2850                   \\ 
 \textcolor{gray}{C$_{6}$H         } & \textcolor{gray}{3880} & \textcolor{gray}{500$^d$              } & \textcolor{gray}{5337} & \textcolor{gray}{H$_{2}$CCC        } & \textcolor{gray}{2110} & \textcolor{gray}{500                     } & \textcolor{gray}{2110} & \textcolor{gray}{P               } & \textcolor{gray}{ 750} & \textcolor{gray}{375                  } & \textcolor{gray}{    } \\ 
 \textcolor{gray}{C$_{6}$H$_{2}$   } & \textcolor{gray}{4130} & \textcolor{gray}{500$^d$              } & \textcolor{gray}{5787} & \textcolor{gray}{H$_{2}$CN         } & \textcolor{gray}{2400} & \textcolor{gray}{500                     } & \textcolor{gray}{2400} & \textcolor{gray}{PH              } & \textcolor{gray}{ 800} & \textcolor{gray}{400                  } & \textcolor{gray}{    } \\ 
 \textcolor{gray}{C$_{6}$H$_{6}$   } & \textcolor{gray}{7590} & \textcolor{gray}{500                  } & \textcolor{gray}{7587} & H$_{2}$CO                            & 3260                   &  \hspace{1 mm} 60$^j$                      & 2050                   & \textcolor{gray}{PH$_{2}$        } & \textcolor{gray}{ 850} & \textcolor{gray}{425                  } & \textcolor{gray}{    } \\ 
 \textcolor{gray}{C$_{7}$          } & \textcolor{gray}{4430} & \textcolor{gray}{500$^d$              } & \textcolor{gray}{5600} & \textcolor{gray}{H$_{2}$CS         } & \textcolor{gray}{2025} & \textcolor{gray}{500$^f$                 } & \textcolor{gray}{2700} & \textcolor{gray}{PN              } & \textcolor{gray}{1100} & \textcolor{gray}{500                  } & \textcolor{gray}{    } \\ 
 \textcolor{gray}{C$_{7}$H         } & \textcolor{gray}{4680} & \textcolor{gray}{500$^d$              } & \textcolor{gray}{6137} & H$_{2}$O                             & 4800                   & 100                                    & 4800                   & \textcolor{gray}{PO              } & \textcolor{gray}{1150} & \textcolor{gray}{500                  } & \textcolor{gray}{    } \\ 
 \textcolor{gray}{C$_{7}$H$_{2}$   } & \textcolor{gray}{4930} & \textcolor{gray}{500$^d$              } & \textcolor{gray}{6587} & H$_{2}$O$_{2}$                       & 6000                   & 100$^c$                                    & 5700                   & \textcolor{gray}{S               } & \textcolor{gray}{ 985} & \textcolor{gray}{495$^f$              } & \textcolor{gray}{1100} \\ 
 \textcolor{gray}{C$_{7}$N         } & \textcolor{gray}{5130} & \textcolor{gray}{500$^d$              } & \textcolor{gray}{6400} & H$_{2}$S                             & 2290                   & \hspace{1 mm} 90$^a$                       & 2743                   & \textcolor{gray}{S$_{2}$         } & \textcolor{gray}{2000} & \textcolor{gray}{500$^d$              } & \textcolor{gray}{2200} \\ 
 \textcolor{gray}{C$_{8}$          } & \textcolor{gray}{4830} & \textcolor{gray}{500$^d$              } & \textcolor{gray}{6400} & \textcolor{gray}{H$_{2}$S$_{2}$    } & \textcolor{gray}{2600} & \textcolor{gray}{500$^d$                 } & \textcolor{gray}{3100} & \textcolor{gray}{Si              } & \textcolor{gray}{2400} & \textcolor{gray}{500$^f$              } & \textcolor{gray}{2700} \\ 
 \textcolor{gray}{C$_{8}$H         } & \textcolor{gray}{5080} & \textcolor{gray}{500$^d$              } & \textcolor{gray}{6937} & \textcolor{gray}{H$_{2}$SiO        } & \textcolor{gray}{1200} & \textcolor{gray}{500                     } & \textcolor{gray}{1200} & \textcolor{gray}{SiC             } & \textcolor{gray}{3150} & \textcolor{gray}{500$^f$              } & \textcolor{gray}{3500} \\ 
 \textcolor{gray}{C$_{8}$H$_{2}$   } & \textcolor{gray}{5340} & \textcolor{gray}{500$^d$              } & \textcolor{gray}{7387} & \textcolor{gray}{HC$_{2}$O         } & \textcolor{gray}{2010} & \textcolor{gray}{500$^d$                 } & \textcolor{gray}{2400} & \textcolor{gray}{SiC$_{2}$       } & \textcolor{gray}{1300} & \textcolor{gray}{500                  } & \textcolor{gray}{1300} \\ 
 \textcolor{gray}{C$_{9}$          } & \textcolor{gray}{5640} & \textcolor{gray}{500$^d$              } & \textcolor{gray}{7200} & \textcolor{gray}{HC$_{2}$P         } & \textcolor{gray}{1400} & \textcolor{gray}{500                     } & \textcolor{gray}{    } & \textcolor{gray}{SiC$_{2}$H      } & \textcolor{gray}{1350} & \textcolor{gray}{500                  } & \textcolor{gray}{1350} \\ 
 \textcolor{gray}{C$_{9}$H         } & \textcolor{gray}{5890} & \textcolor{gray}{500$^d$              } & \textcolor{gray}{7737} & \textcolor{gray}{HC$_{3}$N         } & \textcolor{gray}{2685} & \textcolor{gray}{500$^f$                 } & \textcolor{gray}{4580} & \textcolor{gray}{SiC$_{2}$H$_{2}$} & \textcolor{gray}{1400} & \textcolor{gray}{500                  } & \textcolor{gray}{1400} \\ 
 \textcolor{gray}{C$_{9}$H$_{2}$   } & \textcolor{gray}{6140} & \textcolor{gray}{500$^d$              } & \textcolor{gray}{8187} & \textcolor{gray}{HC$_{5}$N         } & \textcolor{gray}{4180} & \textcolor{gray}{500$^d$                 } & \textcolor{gray}{6180} & \textcolor{gray}{SiC$_{3}$       } & \textcolor{gray}{1600} & \textcolor{gray}{500                  } & \textcolor{gray}{    } \\ 
 \textcolor{gray}{C$_{9}$N         } & \textcolor{gray}{6340} & \textcolor{gray}{500$^d$              } & \textcolor{gray}{8000} & \textcolor{gray}{HC$_{7}$N         } & \textcolor{gray}{5390} & \textcolor{gray}{500$^d$                 } & \textcolor{gray}{7780} & \textcolor{gray}{SiC$_{3}$H      } & \textcolor{gray}{1650} & \textcolor{gray}{500                  } & \textcolor{gray}{1650} \\ 
 \textcolor{gray}{C$_{10}$         } & \textcolor{gray}{8000} & \textcolor{gray}{500                  } & \textcolor{gray}{8000} & \textcolor{gray}{HC$_{9}$N         } & \textcolor{gray}{6590} & \textcolor{gray}{500$^d$                 } & \textcolor{gray}{9380} & \textcolor{gray}{SiC$_{4}$       } & \textcolor{gray}{1900} & \textcolor{gray}{500                  } & \textcolor{gray}{1900} \\ 
 \textcolor{gray}{C$_{10}$H        } & \textcolor{gray}{8540} & \textcolor{gray}{500                  } & \textcolor{gray}{    } & \textcolor{gray}{HCCN              } & \textcolor{gray}{2270} & \textcolor{gray}{500$^d$                 } & \textcolor{gray}{3780} & \textcolor{gray}{SiCH$_{2}$      } & \textcolor{gray}{1100} & \textcolor{gray}{500                  } & \textcolor{gray}{1100} \\ 
 \textcolor{gray}{C$_{10}$H$_{2}$  } & \textcolor{gray}{8990} & \textcolor{gray}{500                  } & \textcolor{gray}{    } & \textcolor{gray}{HCN               } & \textcolor{gray}{1580} & \textcolor{gray}{500$^f$                 } & \textcolor{gray}{2050} & \textcolor{gray}{SiCH$_{3}$      } & \textcolor{gray}{1150} & \textcolor{gray}{500                  } & \textcolor{gray}{1150} \\ 
 \textcolor{gray}{C$_{11}$         } & \textcolor{gray}{8800} & \textcolor{gray}{500                  } & \textcolor{gray}{    } & \textcolor{gray}{HCNO              } & \textcolor{gray}{2850} & \textcolor{gray}{500                     } & \textcolor{gray}{    } & \textcolor{gray}{SiH             } & \textcolor{gray}{2620} & \textcolor{gray}{500$^f$              } & \textcolor{gray}{3150} \\ 
 \textcolor{gray}{CCP              } & \textcolor{gray}{1350} & \textcolor{gray}{500                  } & \textcolor{gray}{    } & \textcolor{gray}{HCO               } & \textcolor{gray}{1355} & \textcolor{gray}{500$^f$                 } & \textcolor{gray}{1600} & \textcolor{gray}{SiH$_{2}$       } & \textcolor{gray}{3190} & \textcolor{gray}{500$^d$              } & \textcolor{gray}{3600} \\ 
 \textcolor{gray}{CCl              } & \textcolor{gray}{1150} & \textcolor{gray}{500                  } & \textcolor{gray}{1150} & \textcolor{gray}{HCOOCH$_{3}$      } & \textcolor{gray}{4000} & \textcolor{gray}{500                     } & \textcolor{gray}{4000} & \textcolor{gray}{SiH$_{3}$       } & \textcolor{gray}{3440} & \textcolor{gray}{500$^d$              } & \textcolor{gray}{4050} \\ 
 \textcolor{gray}{CH               } & \textcolor{gray}{ 590} & \textcolor{gray}{295$^f$              } & \textcolor{gray}{ 925} & HCOOH                                & 4532                   & 150$^a$                                    & 5000                   & \textcolor{gray}{SiH$_{4}$       } & \textcolor{gray}{3690} & \textcolor{gray}{500$^d$              } & \textcolor{gray}{4500} \\ 
 \textcolor{gray}{CH$_{2}$         } & \textcolor{gray}{ 860} & \textcolor{gray}{430$^f$              } & \textcolor{gray}{1050} & \textcolor{gray}{HCP               } & \textcolor{gray}{1100} & \textcolor{gray}{500                     } & \textcolor{gray}{    } & \textcolor{gray}{SiN             } & \textcolor{gray}{3500} & \textcolor{gray}{500                  } & \textcolor{gray}{3500} \\ 
 \textcolor{gray}{CH$_{2}$CCH      } & \textcolor{gray}{3840} & \textcolor{gray}{500                  } & \textcolor{gray}{    } & \textcolor{gray}{HCS               } & \textcolor{gray}{2000} & \textcolor{gray}{500$^d$                 } & \textcolor{gray}{2350} & \textcolor{gray}{SiNC            } & \textcolor{gray}{1350} & \textcolor{gray}{500                  } & \textcolor{gray}{1350} \\ 
 \textcolor{gray}{CH$_{2}$CCH$_{2}$} & \textcolor{gray}{4290} & \textcolor{gray}{500                  } & \textcolor{gray}{    } & \textcolor{gray}{HCSi              } & \textcolor{gray}{1050} & \textcolor{gray}{500                     } & \textcolor{gray}{1050} & \textcolor{gray}{SiO             } & \textcolor{gray}{3150} & \textcolor{gray}{500$^f$              } & \textcolor{gray}{3500} \\ 
 \textcolor{gray}{CH$_{2}$CHCCH    } & \textcolor{gray}{5090} & \textcolor{gray}{500                  } & \textcolor{gray}{    } & \textcolor{gray}{He                } & \textcolor{gray}{ 100} & \textcolor{gray}{{\hspace{0.1 mm} 50}$^d$} & \textcolor{gray}{ 100} & \textcolor{gray}{SiO$_{2}$       } & \textcolor{gray}{4300} & \textcolor{gray}{500                  } & \textcolor{gray}{4300} \\ 
 \textcolor{gray}{CH$_{2}$CHCN     } & \textcolor{gray}{5480} & \textcolor{gray}{500                  } & \textcolor{gray}{5480} & \textcolor{gray}{HF                } & \textcolor{gray}{ 500} & \textcolor{gray}{250                     } & \textcolor{gray}{    } & \textcolor{gray}{SiS             } & \textcolor{gray}{3400} & \textcolor{gray}{500$^f$              } & \textcolor{gray}{3800} \\ 
 \textcolor{gray}{CH$_{2}$CN       } & \textcolor{gray}{2470} & \textcolor{gray}{500$^d$              } & \textcolor{gray}{4230} & \textcolor{gray}{HNC               } & \textcolor{gray}{1510} & \textcolor{gray}{500$^d$                 } & \textcolor{gray}{2050} & \textcolor{gray}{SO              } & \textcolor{gray}{1800} & \textcolor{gray}{500$^f$              } & \textcolor{gray}{2600} \\ 
 \textcolor{gray}{CH$_{2}$CO       } & \textcolor{gray}{2520} & \textcolor{gray}{500$^d$              } & \textcolor{gray}{2200} & \textcolor{gray}{HNC$_{3}$         } & \textcolor{gray}{4580} & \textcolor{gray}{500                     } & \textcolor{gray}{4580} & SO$_{2}$                           & 3010                   & 110$^a$                                 & 5330                   \\ 
 \textcolor{gray}{CH$_{2}$NH       } & \textcolor{gray}{1560} & \textcolor{gray}{500                  } & \textcolor{gray}{3428} & \textcolor{gray}{HNCHO             } & \textcolor{gray}{3980} & \textcolor{gray}{500                     } & \textcolor{gray}{    } & \textcolor{gray}{                } & \multicolumn{2}{c}{  }                                                                    \\
 \textcolor{gray}{CH$_{2}$OH       } & \textcolor{gray}{2170} & \textcolor{gray}{500$^d$              } & \textcolor{gray}{5084} & \textcolor{gray}{HNCO              } & \textcolor{gray}{2270} & \textcolor{gray}{500$^d$                 } & \textcolor{gray}{2850} & \textcolor{gray}{                } & \multicolumn{2}{c}{  }                                                                    \\
 \hline
 \multicolumn{12}{p{15cm}}{References: $^a$Estimated from \cite{Collings:2004}; 
                          $^b$\cite{Smith:2016} with change in pre-factor;
                          $^c$\cite{Dulieu:2013}; 
                          $^d$\cite{Hasegawa:1993A}; 
                          $^e$Similar to \cite{Acharyya:2014}; 
                          $^f$Average between \cite{Hasegawa:1993A} and \cite{Aikawa:1996} values; 
                          $^h$Average between \cite{Tielens:1987} and \cite{Aikawa:1996} values;
                          $^i$\cite{He:2014,He:2014A}; 
                          $^j$\cite{Noble:2012a}; 
                          $^k$\cite{Al-Halabi:2007}}
                     
 \label{energies}
 \end{longtable}
 
\section{Chemical model} \label{sec:method}
We have simulated the time-dependent gas-grain chemistry of an homogeneous dark cloud with 
constant
physical conditions. This is a two-phase chemistry model, since it treats the gas 
and solid phase; however, without using location information of ice species within the ice 
mantle. The model, therefore, does not 
differentiate between reactions between species located on the ice surface and in the bulk 
mantle; however, grain-surface reactions 
are limited to occur within two monolayers worth of material in the ice mantle only. 
Here we give a brief explanation of the model. 
For more details we refer to \cite{Walsh:2015} and references therein. 

\subsection{Gas-grain network and physical conditions}
The gas-phase network is the UDfA Database for
Astrochemistry \citep{McElroy:2013}, known as \textsc{Rate12}. This network does not 
account for three-body reactions, since these are not important at the density used here. 
Photoreactions and direct cosmic ray ionization are included. The cosmic 
ray ionization rate ($\zeta$) used is 1.3~$\times$~10$^{-17}$~s$^{-1}$ \citep{Indriolo:2015}. 

The solid phase chemistry is based on the Ohio State University (OSU)
network\footnote{http://faculty.virginia.edu/ericherb/} \citep{Garrod:2008b}, 
which includes gas-grain interactions such as desorption and adsorption processes,
and grain surface chemistry. We neglect cosmic-ray-induced thermal desorption and reactive desorption 
because the rates for these processes remain very uncertain compared with thermal desorption and photodesorption.
The model also includes grain-cation recombination. We assume a spherical and compact 
grain with radius of 0.1~$\mu$m for simplicity. The grain abundance is fixed to 
1.3~$\times$~10$^{-12}$ with respect to H nuclei, and the density of grain surface
sites is 1.5~$\times$~10$^{15}$~cm$^{-2}$. The ratio between the diffusion barrier 
($E\rm_{diff}$) and the molecular binding energy is assumed to 
be $E\rm_{diff}$~=~0.3 $\times$ $E\rm_{bind}$. The value of $E\rm_{diff}$/$E\rm_{bind}$ is still under debate, 
and most modelers \citep{Hasegawa:1992, Ruffle:2000, Garrod:2006a, Cuppen:2009} have used values 
ranging from 0.3 to 0.8, although the consensus is that for stable species this ratio should be 
between 0.3 and 0.4 \citep{Karssemeijer:2014III}. Here we chose 0.3, which is an optimistic value that 
allows the radicals in the grain mantle to diffuse with some efficiency at low temperatures.

In the subsequent sections, we present results for many thousands of simulations for a fixed set of physical conditions and 
input parameters, except for the set of molecular binding energies. The physical conditions 
used here are those typical for a dark cloud. The initial abundances are taken from \cite{McElroy:2013}, 
which are identical to those used in \cite{Garrod:2009}. These initial abundances follow the lower metallicity set from 
\cite{Graedel:1982}. The total H nuclei density is 2~$\times$~10$^{4}$~cm$^{-3}$, and the
visual extinction ($A\rm_{V}$) is 10~mag. Both gas and dust temperatures are 10 K. 
All simulations were performed over a timescale of 10$^{8}$~years, at which time steady 
state is expected to be reached. Table~\ref{initial_abundances} summarizes the initial elemental 
abundances, while Table~\ref{parameters} summarizes the assumed physical parameters. This is a 0D model, 
which means that all physical parameters are kept constant during the simulations.

In each simulation, a 
value of the binding energy for each grain-surface species is chosen at random from 
a normal probability distribution. The mean value of the normal distribution is the
binding energy and the uncertainties correspond to a 3 $\sigma$ error. The diffusion rates 
for each species $i$ are than calculated using

\begin{equation}
 k_{{\rm thermal,}i} = \nu~ {\rm exp}\Big(-0.3\frac{E_{\rm{bind},i}}{k_{{\rm B}}T}\Big)
\end{equation}

\noindent and

\begin{equation}
 k_{{\rm quantum,}i} = \nu~ {\rm exp}\Big(-\frac{2a}{\hbar}\sqrt{0.6\mu E_{{\rm bind,}i}}\Big),
\end{equation}

\noindent by assuming thermal or quantum processes. Quantum diffusion is only allowed for light species such as H and H$_{2}$. 
In the equations above, $\nu$, $a$, and $\mu$ are 
the frequency, the barrier width, and the reduced mass, respectively. These equations show that 
the binding energies have strong influence on the diffusion rates. Typical values for $\nu$ are 10$^{12}$ s$^{-1}$ and we adopt a 
barrier width between surface sites of 1.5 \AA. 

\begin{table}
\caption{Initial abundances with respect to H nuclei.} \label{initial_abundances}
 \begin{center}
  \begin{tabular}{cc}
  \hline
  \hline
  Species & \multicolumn{1}{c}{Abundances} \\
  \hline
  H             & 5.00(-05)  \\
  H$_{2}$       & 5.00(-01)  \\
  He            & 9.75(-02)  \\
  C             & 1.40(-04)  \\
  N             & 7.50(-05)  \\
  O             & 3.20(-04)  \\
  F             & 2.00(-08)  \\
  Na            & 2.00(-09)  \\
  Mg            & 7.00(-09)  \\
  Si            & 8.00(-09)  \\
  P             & 3.00(-09)  \\
  S             & 8.00(-08)  \\
  Cl            & 4.00(-09)  \\
  Fe            & 3.00(-09)  \\
  Grain density & 1.30(-12)  \\
  \hline
  \multicolumn{2}{c}{NOTE. - The notation $\alpha$($\beta$) stands for $\alpha\times$10$^{\beta}$}
  \end{tabular} \\
 \end{center}
\end{table}

\begin{table}
\caption{Dark cloud physical parameters.}
 \begin{center}
  \begin{tabular}{c c}
  \hline
  \hline
  $n\rm_{H}$    & 2$\times$10$^{4}$ cm$^{-3}$ \\
  $T\rm_{dust}$ & 10 K     \\
  $T\rm_{gas}$  & 10 K     \\ 
  $A\rm_{V}$    & 10 mag   \\
  $\zeta$       & 1.3 $\times$ 10$^{-17}$ s$^{-1}$ \\
  \hline
  \end{tabular} \\
 \end{center}
 \label{parameters}
\end{table}

\section{Results} \label{section:results}
\subsection{Updating the network} \label{section:updatenetwork}
The first set of models used the standard network from \cite{Drozdovskaya:2014, Drozdovskaya:2015}, and \cite{Walsh:2015}, 
which adopt the surface network from \cite{Garrod:2008b} and mainly focussed on O-bearing complex organic molecules.
We ran a thousand simulations using the standard network.  For the key ice species, CO, HCO, 
H$_{2}$CO, CH$_{3}$OH, HCOOH, H$_{2}$O, CO$_{2}$, CH$_{4}$, and NH$_{3}$, we then calculated 
the Pearson correlation coefficient between the logarithm of the abundance at different times 
and the binding energy of each of the species listed in Table~\ref{energies}

\begin{equation}
 P(A,B) = \frac{\text{cov}(\log(n(\text{A})),E_{\text{bind}}(\text{B}))}{ \sigma (\log(n(\text{A}))) \sigma (E_{\text{bind}}(\text{B}))},
\end{equation}

\noindent where cov stands for the covariance and $\sigma$ is the standard deviation. This coefficient
gives a measure of linear correlation or anti-correlation between the logarithm of the
ice abundance and the binding energy, and lies between -1 and 1. We selected those species
that have an absolute correlation coefficient equal to or larger than 0.3, i.e. $|P|$~$\geq$~0.3. 
The binding energy of HNO was found to correlate with the log of the abundance of several species,
which is rather unexpected since HNO is not a direct precursor to any of these species.
HNO acts, however, as an efficient producer of OH radicals via the reaction

\begin{equation}
 \text{HNO} + \text{O} \rightarrow \text{NO} + \text{OH}.
\end{equation}

\noindent HNO is formed through hydrogenation of NO. A dominant destruction reaction for
HNO in our standard network is the reaction with H leading to NO and H$_{2}$. NO catalyzes in 
this way the production of OH from O and H by making H temporarily unavailable for other 
reactions. However, laboratory experiments \citep{Congiu:2012,Fedoseev:2012} show that hydrogenation of HNO leads predominantly 
to the formation of NH$_{2}$OH through

\begin{equation}
 \text{HNO} + \text{H} \rightarrow \text{H}_2\text{NO}
\end{equation}

\noindent followed by

\begin{equation}
 \text{H}_2\text{NO} + \text{H} \rightarrow \text{NH}_2\text{OH}.
\end{equation}

These reactions will take HNO out of the loop and block the formation of new OH
radicals that can lead to the formation of CH$_{3}$OH, HCOOH, CO$_{2}$, and H$_{2}$O. We have
therefore updated the network by adding the chemistry for NH$_{2}$OH, which includes
the gas phase chemistry of the neutrals and all intermediate ions. We have also included
the chemistry for NH$_{2}$CHO (formamide), and CH$_{3}$NH$_{2}$ (methylamine), both gas and
surface reactions and also extracted from the OSU network \citep{Garrod:2008b}.
This first set of simulations shows clearly the importance of new laboratory 
experiments to investigate chemical pathways that are not yet considered in most of astrochemical 
models but are fundamental to obtain more reliable results. The final network contains 8971 reactions connected by 689 species.

\subsection{Varying binding energies} \label{section:energies}
\begin{figure}
\begin{center}
  \centering
  \includegraphics[width=\textwidth]{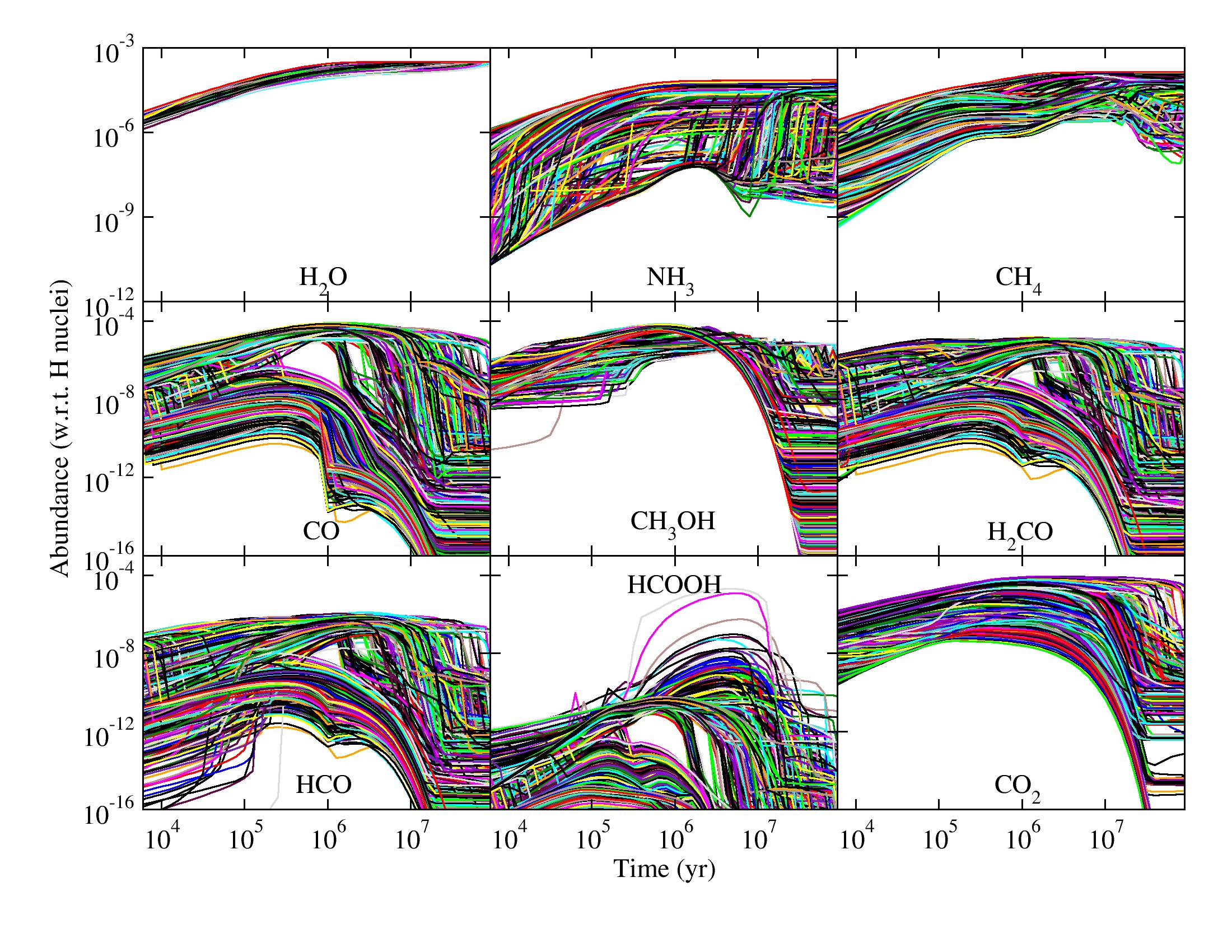}
  \caption{Evolution of the ice abundance of important species. Each panel shows 10,000 simulations, where the 
   binding energies used were randomly chosen from a normal distribution. All abundances are shown with respect to H nuclei. 
   Abundances much lower than 10$^{-16}$ are negligible and are therefore not shown here.}
     \label{evolution}
\end{center}
\end{figure}

\begin{figure}
\begin{center}
  \centering
  \includegraphics[width=\textwidth]{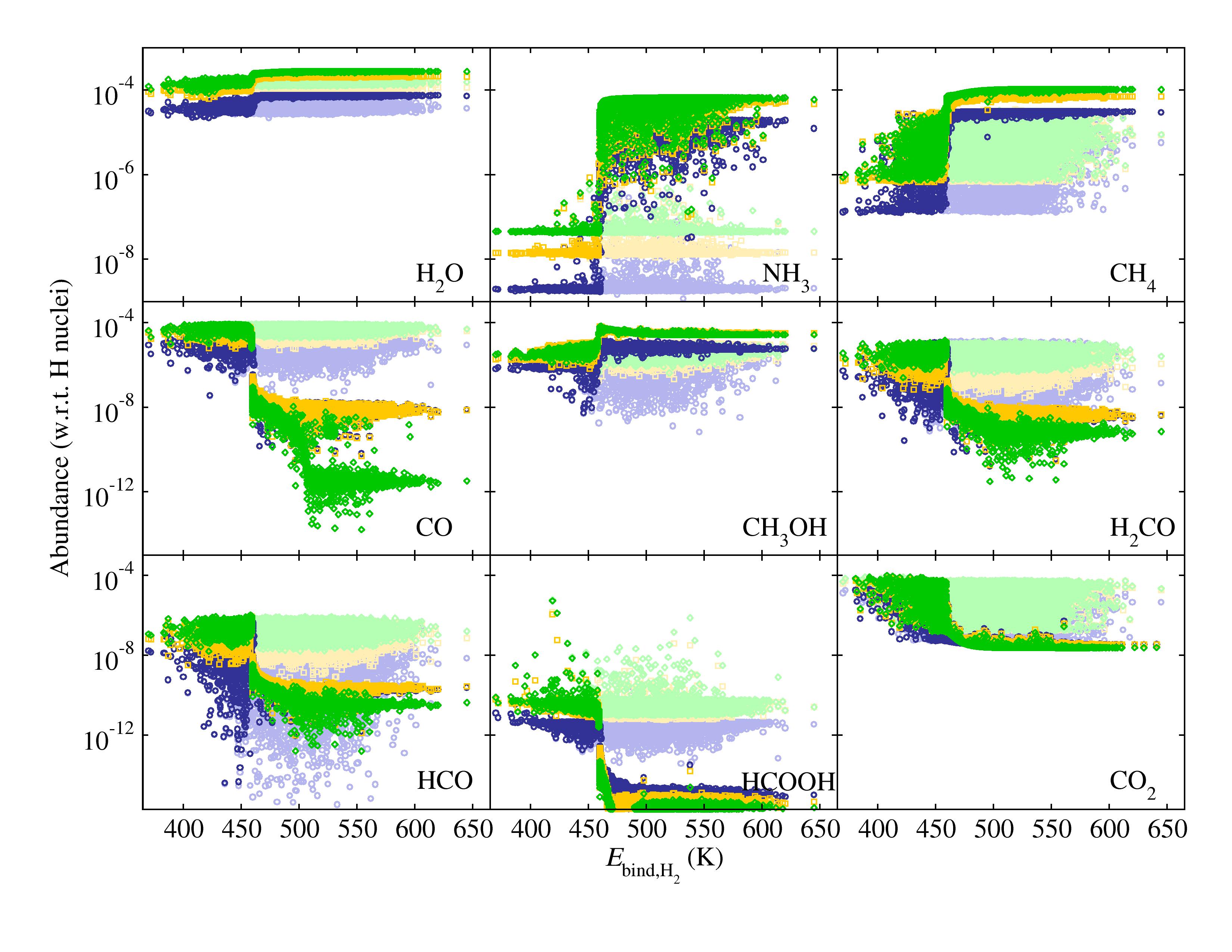}
  \caption{Abundance of ice species as a function of H$_{2}$ binding energy. Each panel shows abundances from 
   10,000 simulations, each one derived from a different set of binding energies. Results for 10$^{5}$ years are shown in blue, while 
   in yellow are shown results for 5$\times$10$^{5}$ years, and for 10$^{6}$ years in green. All abundances are calculated with 
   respect to H nuclei. Abundances much lower than 10$^{-16}$ are negligible and are therefore not shown here. 
   The darker colors represent the original model, whereas the lighter colors represent results using a model that reduces H$_2$ freeze out.}
     \label{H2_correlation}
\end{center}
\end{figure}

Using this new network we performed 10,000 simulations selecting a set of binding energies at random from their normal 
distributions. Figure~\ref{evolution} shows the 
evolution of the ice abundance of some key ice species~--~CO, HCO, H$_{2}$CO, CH$_{3}$OH, HCOOH, H$_{2}$O, CO$_{2}$, CH$_{4}$, 
and NH$_{3}$. The abundance is taken with respect to H nuclei. Clearly there is a large variation 
in steady state abundance, especially for the species with CO as one of their precursors, 
but also in the time evolution of the abundances. To assess the origin of these differences, 
we again determined the Pearson correlation coefficient between the binding energies and the 
log of the abundance with respect to H nuclei. The binding energy of H$_{2}$ was found to be the most 
strongly correlating parameter. This is evident in Figure~\ref{H2_correlation}, which shows the ice abundance at three 
different times~--~10$^{5}$ (darker blue), 5$\times$10$^{5}$ (darker yellow), and 10$^{6}$ (darker green) years~--~as a function 
of H$_{2}$ binding energy. There is a clear change in chemistry around $E\rm_{bind,H_{2}}$~=~460~K. For some species this 
change occurs somewhat earlier than for others. Two groups can be identified according to similar
general behavior. A group, including CH$_{4}$ and NH$_{3}$, that have a high abundance for high $E\rm_{bind,H_{2}}$ 
and a group with CO, HCO, H$_{2}$CO, CO$_{2}$, and HCOOH which show a sharp decrease 
with increasing $E\rm_{bind,H_{2}}$. The abundance of water and methanol is less sensitive to the value of  H$_{2}$ binding energy.

To determine the origin of this bifurcation, we have analyzed the two populations containing runs with
$E\rm_{bind,H_{2}} < $~460~K and $E\rm_{bind,H_{2}} > $ 465 K separately. By identifying 
correlations between ice abundances 
and binding energies in both simulation populations separately, we hope to gain insight into 
the change in chemistry that is causing this effect. The discontinuity in temperature is 
intentional in order to isolate the two sets of results. The binding energy 
of H$_{2}$ still correlates most frequently with the different ice abundances, mainly at 
later times. Other species 
whose binding energies appear to correlate strongly are HCO, H, N, CH$_{2}$, and HNO.


\begin{figure}
\begin{center}
  \centering
  \includegraphics[width=0.5\textwidth]{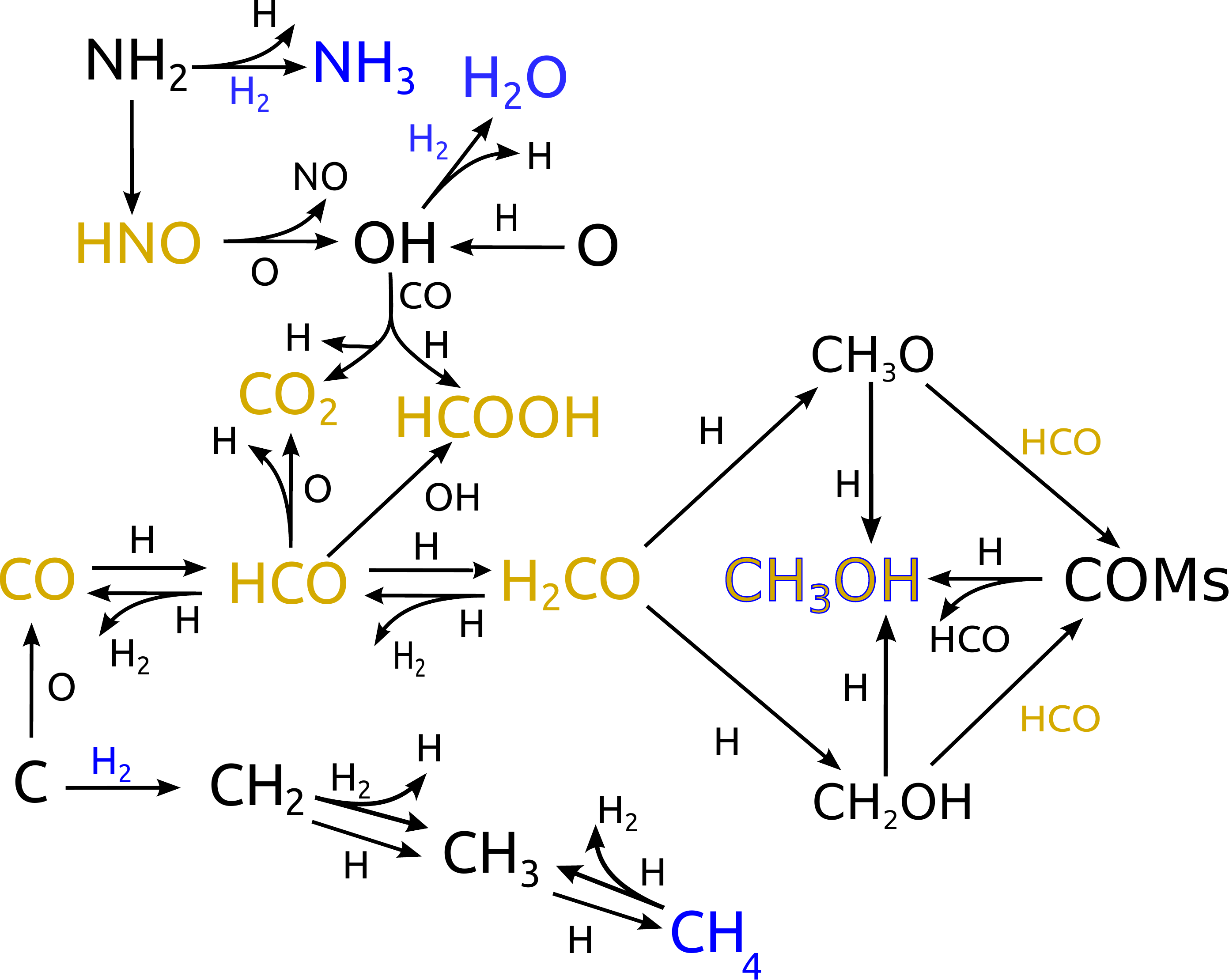}
   \caption{A limited surface network explaining the very different behavior between models with $E\rm_{bind,H_{2}} > $~465~K (indicated 
   in blue) and those with $E\rm_{bind,H_{2}} < $ 460 K (indicated in yellow).}
     \label{network}
\end{center}
\end{figure}  

Using these correlations together with information on the most important formation and 
destruction reactions for the ice species of interest, we extract the limited reaction 
network that explains the observed trends in the results from the full reaction network 
containing 2164 surface reactions. This network is depicted in Figure~\ref{network}. Important 
precursors in this network are NH$_{2}$ (formed from N), and C. With increasing H$_{2}$ binding energies, the 
residence time of H$_{2}$ on the surface increases whereas the diffusion barrier for all cases is low enough for H$_{2}$ to be 
mobile. For a sufficiently high $E\rm_{bind,H_{2}}$~C and NH$_{2}$ react sequentially with H$_{2}$ to form CH$_{4}$ and NH$_{3}$, respectively.  
These pathways are indicated in blue in the schematic. Water is predominantly made through
\begin{equation}
 \text{OH} + \text{H}_{2} \rightarrow \text{H}_2\text{O} + \text{H}
\end{equation}
and its abundance therefore increases with H$_{2}$ binding energy. The hydroxyl radical,
OH, can be formed in different ways. We found the dominant channel to be H + O. For low H$_{2}$ binding energies,
\begin{equation}
 \text{O} + \text{HCO} \rightarrow \text{CO} + \text{OH}
\end{equation}
and
\begin{equation}
 \text{H} + \text{H}_{2}\text{O}_{2} \rightarrow \text{H}_{2}\text{O} + \text{OH}
\end{equation}
become important as well. 

At some critical H$_{2}$ binding energy, the H$_{2}$ residence time becomes too short to be able to compete 
with oxidation reactions of both C and NH$_{2}$, leading to the formation of CO and HNO, respectively. 
This opens up a new type of chemistry leading to the species indicated in yellow, CO, HCO, H$_{2}$CO, 
CO$_{2}$, and HCOOH. 

Methanol is not only a precursor for more complex saturated
organic molecules (COMs) but also one of their main destruction products. We find that
for high H$_{2}$ binding energies, CH$_{3}$OH is formed exclusively via hydrogenation of CH$_{3}$O and CH$_{2}$OH, which are direct 
products of H$_{2}$CO~+~H, even though most of the carbon is initially converted to CH$_4$. For low H$_{2}$ binding energies, however, 
the ice abundance of HCO increases and CH$_{3}$O and CH$_{2}$OH are found to react with HCO instead of H leading to more complex species. 
In this way, methanol is more or less ``skipped'' in the reaction network and is predominantly formed through the  destruction of more 
complex species (i.e., there opens up a ``top-down'' interstellar formation route to methanol ice).

For $E\rm_{bind,H_{2}} > 465$~K, the H$_2$ surface abundance was found to be equivalent 
to a thick H$_2$ ice. This is unrealistic to occur, since the binding energy of H$_2$ to 
itself is significantly lower than to the grain surface. For high $E\rm_{bind,H_{2}}$, a 
maximum of one monolayer of H$_2$ is expected to cover the surface of grain mantle. Heavier 
species landing on the grain will go through this layer to adsorb. H$_{2}$ molecules landing 
or diffusing over another H$_2$ molecule will desorb. We have changed our rate equation model 
to capture this behavior, roughly following \citet{Hincelin:2015}, as detailed in the appendix. 
The correlation with $E\rm_{bind,H_{2}}$ is now found to disappear as can be seen in 
Figure~\ref{H2_correlation} in the lighter colors. In the remainder of the paper we will 
continue with these results, where the H$_2$ surface abundance remains below one monolayer.

\begin{table}
\caption{Species with Pearson correlation coefficients $|P|$ $\geq$ 0.3 and their 
correspondent values. Results are shown for 10$^{5}$, 5$\times$ 10$^{5}$, and 10$^{6}$ years.}
 \begin{center}
  \begin{tabular}{c||lr|lr|lr}
  \hline
  \hline
   & \multicolumn{2}{c}{1$\times$10$^{5}$ yr} & \multicolumn{2}{c}{5$\times$10$^{5}$ yr} & \multicolumn{2}{c}{1$\times$10$^{6}$~yr} \\
\hline
Species    & corr.    &  $P$  & corr.    &  $P$  & corr.    &  $P$  \\
\hline
CH$_{3}$OH & HNO      &  0.39 & CH$_{2}$ &  0.52 & CH$_{2}$ &  0.50 \\
           & CH$_2$   &  0.37 &          &       & HCO      &  0.38 \\

CO         & HCO      &  0.50 & HCO      &  0.83 & HCO      &  0.85 \\
           & CH$_2$   & -0.33 &          &       &          &       \\

CO$_{2}$   & CH$_2$   & -0.56 & HCO      & -0.85 & HCO      & -0.87 \\
           & HNO      &  0.39 &          &       &          &       \\
           & C        & -0.31 &          &       &          &       \\

H$_{2}$O   & HNO      & -0.66 & HCO      &  0.55 & HCO      &  0.65 \\
           &          &       & HNO      & -0.52 & HNO      & -0.41 \\
                              
CH$_{4}$   & C        &  0.66 & C        &  0.62 & C        &  0.57 \\
           & HCO      & -0.33 & HCO      & -0.38 & HCO      & -0.43 \\
           &          &       & HNO      & -0.30 & HNO      &  0.34 \\
                              
HCO        & HCO      &  0.61 & HCO      &  0.94 & HCO      &  0.91 \\
           & CH$_2$   & -0.33 &          &       &          &       \\

H$_{2}$CO  & HCO      &  0.78 & HCO      &  0.93 & HCO      &  0.89 \\
           & C        & -0.36 &          &       &          &       \\

NH$_{3}$   &          &       &          &       &          &       \\
                            
HCOOH      & HCO      &  0.55 & HCO      & 0.66  & HCO      & 0.32   \\
           & C        &  0.31 &          &       &          &       \\

  \hline
  \end{tabular} \\
 \end{center}
 \label{correlation}
\end{table}

\begin{figure}
  \begin{center}
     \centering
     \includegraphics[width=0.8\textwidth]{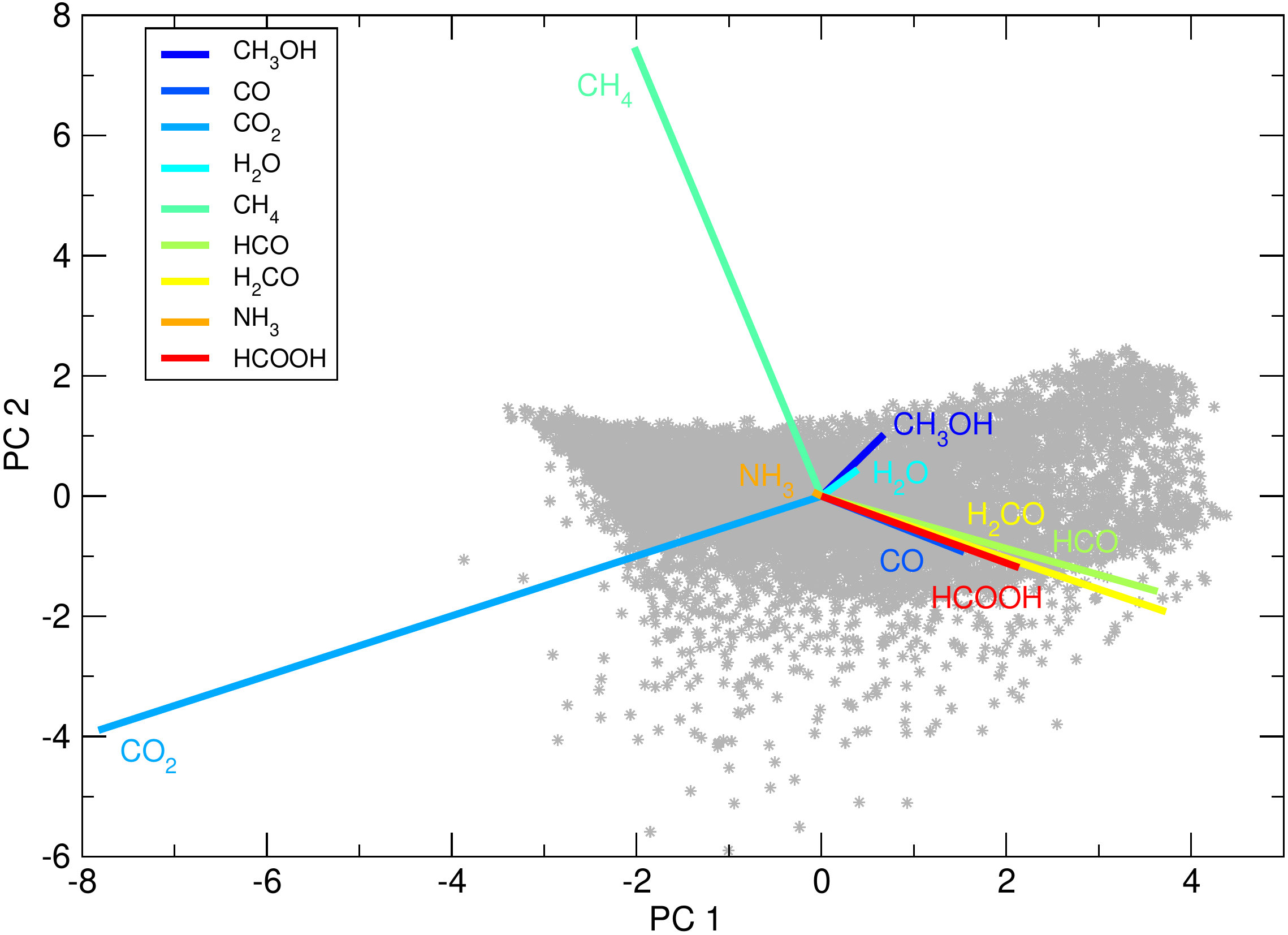}
      \caption{Principal component analysis of the sample of abundances for the main solid species obtained using low H$_{2}$ binding 
      energies. The distribution of abundances over the 
      first and second principal components are shown in gray. The size of the bars represents the dispersion of abundances of each species, 
      where the angle among the bars represents the strength of the correlation between the abundances of these species.}
        \label{PCA_low}
  \end{center}
\end{figure}

The dispersion found in Figure~\ref{H2_correlation} is due to correlations with binding 
energies of species other than H$_2$. This is quantified again by the Pearson correlation 
function for a selection of ice species. All correlations with $|P|\geq 0.3$ are listed in 
Table~\ref{correlation}. As can be seen in this table, CO$_2$ shows a strong correlation 
with the HCO binding energy at late times because an important formation route is through 
HCO~+~O. At early times, the correlation with C and CH$_2$ can be explained by the formation 
of CO through C + O and its competing reaction C + H$_2$ (see Figure~\ref{network}). HNO is 
again linked to the formation of OH which is involved in CO$_2$ production through CO + OH. 
Figure~\ref{H2_correlation} also clearly shows the orders of magnitude difference between 
the ice abundances. To make the correlation between different abundances more apparent, 
we have performed a Principal Component Analysis (PCA) 
of the logarithm of the abundances of the main species obtained at 10$^{6}$ years. In a 
PCA analysis an axis transformation is performed 
where the new dimensions, principal components, are linear combinations of the original 
dimensions; in this case the log of the ice 
abundances. These principal components are chosen such that the first describes as much 
as the variance in the data as is possible, followed by 
the second, etc. The results of the PCA analysis are shown 
in Figure~\ref{PCA_low}. Each gray dot in this figure represents an individual simulation 
projected onto the first and second principal 
components (the scores). The spread of these points show that the dependence of the logarithm 
of the abundance on the parameter choice is highly nonlinear. The lines in the graph represent 
the loadings, which are the coefficients in the transformation matrix. The length 
of the line is a measure for the amount of variance in the original dimension. The length of 
the lines were multiplied by a factor 
of ten for better visualization. In accordance with Figure~\ref{H2_correlation}, the species 
that present the largest variance in 
abundance are CO$_{2}$ and CH$_{4}$, followed by H$_{2}$CO, HCO, HCOOH, CO, HCOOH, CH$_3$OH, 
H$_{2}$O, and NH$_{3}$, respectively. The directions 
of the lines reveal correlations between the abundances, where overlapping lines indicate species 
with similar variance patterns. Although it is dangerous to directly infer a chemical network 
from these correlates, we can however draw some conclusions, on the basis of the PC analysis in 
combination with the flux analysis that we performed earlier to arrive at Figure~\ref{network}.
We see that the CO hydrogenation products, HCO, HCOOH, H$_{2}$CO, as well as CO are correlated. 
CO$_{2}$ and CH$_{3}$OH on the 
other hand do not correlate so tightly indicating that other routes are involved in their 
formation. CO$_{2}$ nearly anticorrelates 
with H$_{2}$O, reflecting a similar precursor, OH. CH$_{4}$ and NH$_{3}$ do not show a 
strong correlation with any of the other 
species, reflecting their rather separate formation routes. The PCA analysis hence confirms 
the scheme depicted in Figure~\ref{network}.

\subsection{The HCO radical and the importance of branching ratios}
\label{ratios}

\begin{figure}
\begin{center}
  \centering
  \includegraphics[width=\textwidth]{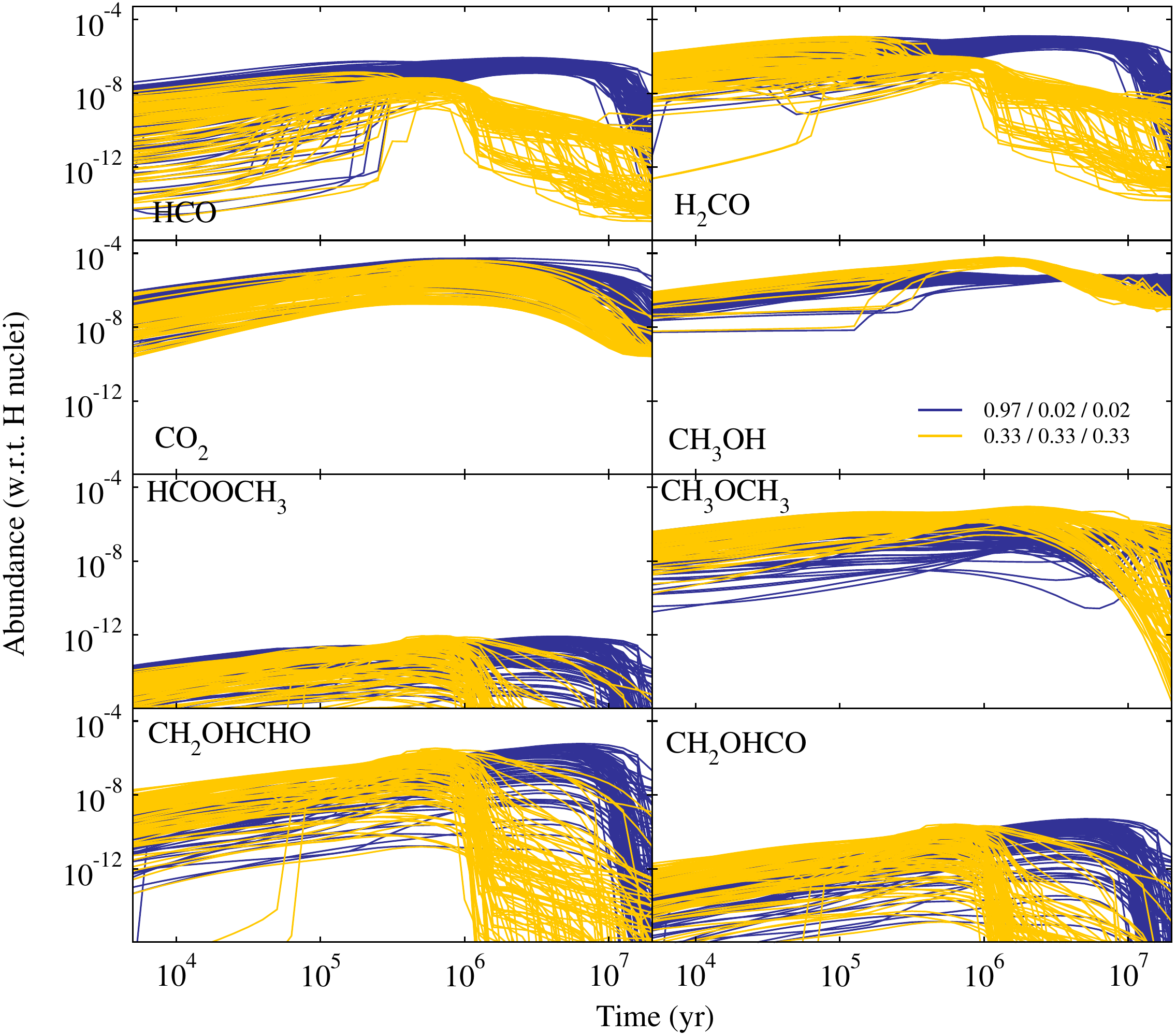}
   \caption{Evolution of ice abundances according to two different branching ratios: 0.97~/~0.02~/~0.02 
   (standard network) and 0.33 / 0.33 / 0.33 (updated network). 
   Results using the standard network are shown in blue and using the updated network in yellow.
   Abundances much lower than 10$^{-16}$ are negligible and are therefore not shown here.}
     \label{diff_rates}
\end{center}
\end{figure}

The Pearson correlation coefficients and the network depicted in Figure~\ref{network} show that HCO plays a pivotal role the grain surface 
network. Here we will look at its formation and destruction routes more closely. Its main formation routes are through hydrogenation of 
CO and H$_2$CO; destruction also mainly occurs through hydrogenation of H$_2$CO. The reaction H + H$_2$CO has three product channels: 
formation of CH$_{3}$O or CH$_{2}$OH and abstraction of a hydrogen atom leading to HCO and H$_{2}$. Using the standard network, the 
combination of reaction rates and branching ratios leads to an effective branching ratio of 0.97 for the latter channel. The remainder 
is equally shared between CH$_{3}$O or CH$_{2}$OH. Whereas this might reflect the gas phase chemistry, we find, however, no laboratory 
data to support these extreme branching ratios on the surface. In a recent experimental study \citet{Chuang:2016} investigated these 
reactions, but they do not quote branching ratios. They find that indeed the addition reaction to CH$_3$O and the abstraction to 
H$_2$ + HCO occur, and they cannot exclude the channel to form CH$_2$OH. Here, we alter the network by adopting equal ratios for all 
three channels and performed 100 additional simulations using this network. The evolution of the 
ice abundance of the species involved in the hydrogenation of CO is shown in Figure~\ref{diff_rates} in yellow. A comparison 
is made between the original 100 runs of our standard network in blue.

The timescale of conversion of HCO to H$_2$CO and CH$_3$OH is reduced by a few million years, 
since there are less back reactions. Moreover, as a consequence of the reduction in HCO, an increase in the methanol abundance can be observed, 
since CH$_{3}$O and CH$_{2}$OH will now predominantly react with H to form methanol. For the more complex species, like HCOOCH$_3$, 
CH$_2$OHCHO, and the intermediate radical CH$_2$OHCO, both a reduction in the peak intensity and a change in the 
formation timescale can be observed. 

This example clearly shows the importance of accurate branching ratios. In this case, for grain-surface reactions with multiple product channels 
the ratio between the forwards and backwards reactions 
is crucial since this determines the timescale of CH$_{3}$OH formation and the route of complex molecule formation: through CH$_3$OH or 
through reactions with HCO. We conclude that, using the standard network, reactions forming complex molecules involving HCO are likely overexpressed. 
Numerous other reactions in the grain surface network have several reaction channels and we believe that the branching ratios used should 
be carefully scrutinized as it is possible that they have strong effects.
   
\subsection{The effect of initial conditions on COMs production}
\label{initial}
The overall results indicate that the C + O $\rightarrow$ CO reaction is crucial in the 
formation of COMs as depicted in Figure~\ref{network}. So far, the initial form of elemental 
carbon in the simulations is chosen to be atomic carbon. However, depending on the history of the molecular cloud a 
substantial fraction could also already be in the form of CO. In this case, one would expect the methanol route start earlier. Here we will 
look at the extreme case in which all elemental carbon is initially in the form of CO. 

\begin{figure}
\begin{center}
  \centering
  \includegraphics[width=\textwidth]{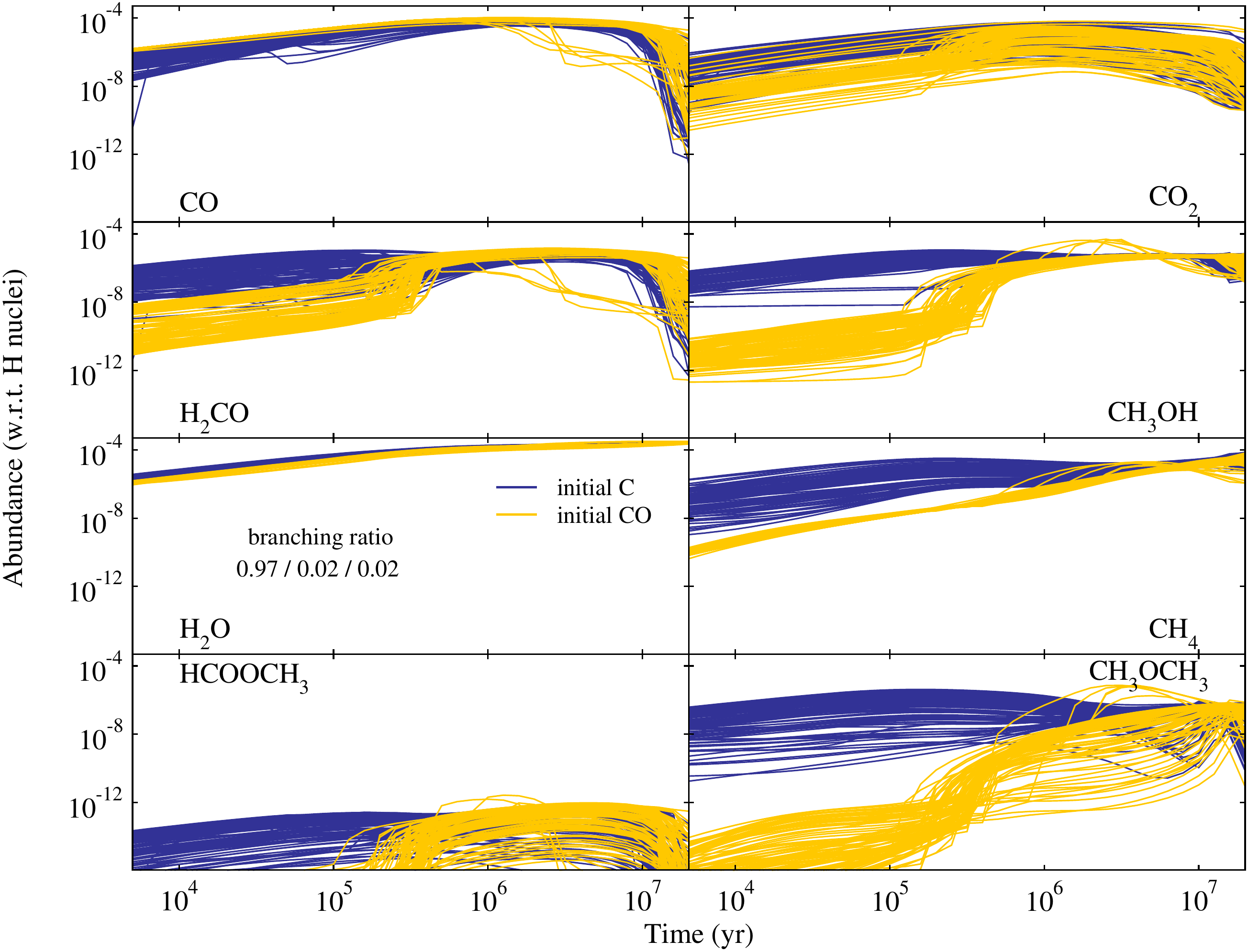}
   \caption{Comparison between the evolution of selected ice phase species using different initial conditions. 
   Dark colors show simulations using the elemental form of carbon as initial conditions, while light colors refer to 
   simulations using CO as the initial form of carbon. Each panel shows 100 simulations for each initial condition, where 
   50 runs use $E\rm_{bind,H_{2}} > $ 465 K (dark and light blue) and other 50 runs using $E\rm_{bind,H_{2}} < $ 460 K (orange 
   and yellow). Here the standard network is used, i.e. the branching ratio of 0.97 / 0.02 / 0.02 for the hydrogenation reactions 
   of H$_{2}$CO. Abundances much lower than 10$^{-16}$ are negligible and are therefore not shown here.}
     \label{COvsC}
\end{center}
\end{figure}

We follow a similar procedure to the previous section by performing another 100 simulations, 
but with all carbon initially in the form of CO. The initial atomic oxygen abundance is reduced accordingly. The standard 
network is used and the results are again compared to the previous results. The results are 
shown in Figure~\ref{COvsC}. These show an unexpected reduction in H$_2$CO, CH$_3$OH and COMs abundance when we 
start with all C in the form of CO. The crucial 
radical to explain this effect is CH$_3$. By starting exclusively from CO, the production rate of this radical is reduced by at least an 
order of magnitude. It can react with for instance HCO to form CH$_3$CHO which can fall back to H$_2$CO upon hydrogenation or lead to more 
complex molecules, which in turn reduce to methanol again. Overall, this careful analysis of the reaction network shows that rather than 
being a final product, COMs also play a crucial role as an intermediate to form main grain mantle species such as formaldehyde and methanol. 
This already occurs in dark conditions and no photoprocesses are needed for COMs to form 
at low temperatures ($\sim$10 K). In hot core models, photodissociation of methanol is used as the formation mechanism of radical 
species that can then recombine to form more complex molecules upon warm up triggered by the birth of the central star 
\citep{Garrod:2006a,Garrod:2008b}. Here we see that they may already form at low temperatures, but that they are destroyed by abstraction reactions 
with atomic hydrogen. The latter reactions become less important for higher temperatures since the residence time of H atoms is significantly 
reduced at temperatures above 20~K. 

\subsection{Comparison with ice observations}
\label{observations}
\begin{figure}
  \begin{center}
     \centering
     \includegraphics[width=0.8\textwidth]{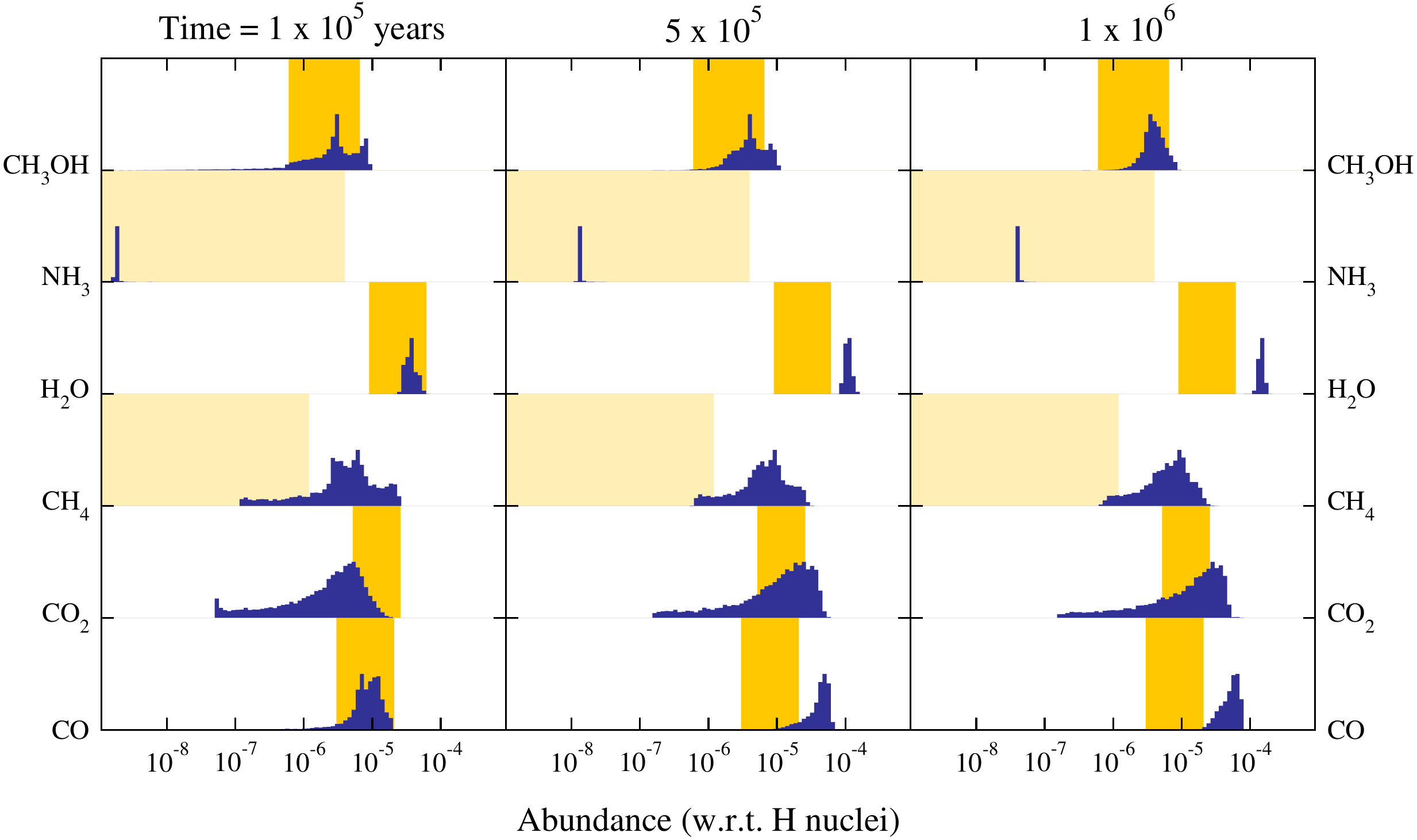}
      \caption{Distributions of simulated ice abundances (blue histograms)  for three different times (10$^5$ in the left, 
      5$\times$10$^5$ in the middle and 10$^6$ years in the right panels) compared to abundances derived from 
      observations (yellow areas, light yellow for upper limits) of quiescent clouds and cores (see \cite{Boogert:2015} for references).}
        \label{comp_obs}
  \end{center}
\end{figure}

Distributions of the obtained ice abundances with
respect to H nuclei are shown in Figure~\ref{comp_obs} for H$_{2}$O, CO, CO$_{2}$, 
CH$_{3}$OH, NH$_{3}$, and CH$_{4}$, which are all securely identified
in these type of objects. Despite the large apparent dispersion in the ice abundances 
in Figure~\ref{H2_correlation}, these histograms highlight that the dispersion in terms 
of full-width-half-maximum is significantly narrower ($\lesssim 1$ order of magnitude) 
for most species. We compare the obtained model abundances with observational abundances 
taken from a recent review by \cite{Boogert:2015} where
we have taken the background star observations as representative of observations of quiescent 
clouds and cores.  What is clear is the very large spread in observational abundances as 
indicated by the yellow areas in Figure~\ref{comp_obs}, from upper limits to relatively large values compared
with water ice observations. Inspection of Figure~\ref{comp_obs} shows that the best agreement is obtained 
at relatively early times of $10^5$ years. This coincides with dark cloud model results 
focusing on the gas phase, where $10^5$ years is considered as the early time of best overall agreement between 
models and observations for a large number of species. The current understanding is that 
the relatively low density of $n_\text{H} = 2 \times 10^4$ cm$^{-3}$ persists until roughly 
1 Myr when the density increases and the remaining gas phase CO freezes out rapidly on the 
grains and the star formation sequence starts. With the current $n_\text{H}$ density CO freeze 
out occurs roughly in  $2\times 10^5$ years (see Figure \ref{COvsC})). After this point, the 
CO adsorption from the gas phase reduces and destruction of the more complex species wins over 
formation, simply because its main precursor has run out.

The observations of CH$_4$ and NH$_3$ show only upper limits. Our model results for NH$_3$ 
fall well below this limit, where as a large fraction of models heavily overproduce CH$_4$. 
This overproduction of CH$_4$ could be due to too efficient destruction reactions of COMs, 
because of inaccuracy in branching ratios of reactions leading to CH$_4$ or its precursors,  
justifying a complete scrutinization of the whole network. 

\section{Discussion and conclusions}
With this paper we set out to study the influence of binding energies of surface species on 
simulated ice abundances in order to constrain some of 
the binding energies. We found that studying the correlations between binding energies and 
simulated ice abundances served as an excellent tool to scrutinize the reaction network. As 
a result we obtained more information on the 
reaction network than actually constraining certain binding energies. One reason for this 
could be the low 
temperature of 10~K applied in the simulations. At these temperatures, most species are 
stationary for their full range of binding energies. 
Hot core or disk models that probe a larger spread in temperature might be more suited for this purpose. However, analyzing the data of a 
simple dark cloud to really understand the dependencies is already very complex and hence serves as a good starting point for similar studies 
of more complex astrophysical environments and we leave this to future work. Finally, we want to stress that only the binding energies were 
varied and that all other parameters are held constant. We expect these to have an effect as well, in particular the diffusion-to-desorption 
ratio. This is also a topic for future study. Moreover, it would be valuable to extend the present work to a more realistic three-phase model, 
so that the bulk and the grain surface can be treated separately. Such a model has be shown to 
capture some of the complexity that arises for the desorption of mixtures of light species such as 
CO, which exhibits multiple desorption peaks \citep{Fayolle:2011}. We arrived at the following 
set of conclusions and recommendations.

\begin{enumerate}
\item For high binding energy of H$_2$, rate equations can result in the unrealistic build  
up of an H$_2$ ice. Rate equations should be corrected for this, since a high surface  
abundance of H$_2$ can trigger a different type of surface chemistry. 
\item For dark cloud models, the binding energies of C, HCO, HNO, and CH$_2$ are the most 
determining for the final ice abundances. 
Since diffusion barriers are inferred from binding energies, these dependencies most likely 
involve diffusion rates. Obtaining accurate 
diffusion rates or binding energies for these species would result in a significant 
improvement in the reliability of grain surface models.
\item HNO is actively involved in the production of OH. Hydrogenation reactions of 
HNO should be included in the network, in accordance with laboratory results, to prevent overproduction of OH.
\item The branching ratios of reaction products resulting from hydrogenation of H$_2$CO 
were found to play a crucial role in determining 
the timescale for COMs formation as well as their formation routes. We believe that with 
the network used here reactions forming complex 
molecules involving HCO are too efficient. We encourage experimentalists to work towards 
extracting branching ratio information from laboratory data:
his work highlights the importance of accurate branching ratios for the outcome of g
ain-surface models.
Since numerous other reactions in the grain surface network have several reaction channels, we recommend modelers on the 
other hand to carefully scrutinize the branching ratios they use as these ratios can have strong effects.
\item COMs  play a crucial role as intermediates for formaldehyde and methanol and 
are not only the final product in the reaction network. 
COMs are thus mainly formed through reactions with HCO, the abundance of which highly depends on the H~+~H$_2$CO branching ratio, 
and subsequently destroyed by abstraction reactions with atomic H. 
\end{enumerate}

\acknowledgments
EMP and HMC acknowledge the European Research Council (ERC-2010-StG, Grant Agreement no. 259510-KISMOL) 
for financial support. HMC is grateful 
for support from the VIDI research program 700.10.427, which is financed by The Netherlands 
Organization for Scientific Research (NWO). 
CW acknowledges support from the VENI research program 639.041.335, also financed by NWO, 
and start-up funds from the University of Leeds. 
We would like to thank Jeroen Jansen (Radboud University) for fruitful discussions about the 
application of PCA and Alexander Atamas for computational support.

\appendix
Rate equations  do not have any positional information of the species. Species in the 
top layer are, in principal, treated in the same way as in the bulk of the ice. This 
results in too many species that actively participate in surface reactions and in the 
wrong desorption order. 
The current rate equation models applies a fix to artificially account for this, as 
explained in \citet{Cuppen:2017}. 
In this treatment, surface reactions do not depend on the number of reactants on the 
grain ($n_\text{s}(\text{A})n_\text{s}(\text{B})$ for reaction A+B$\rightarrow$ C) but 
on the number of reactants in the active layer of size $N_\text{act}$  
($\chi(\text{A})N_\text{act}\chi(\text{B})N_\text{act}$ with $\chi(\text{A}) = n_\text{s}(\text{A})/\sum_X n_\text{s}(X)$). 
Hence, a homogeneous distribution of all species is assumed throughout the grain mantle. 
If the total number of species on the grain $\sum_X n_\text{s}(X)$ is less than  $N_\text{act}$, 
the original expression with is $n_\text{s}(\text{A})n_\text{s}(\text{B})$ is used.

Molecular hydrogen is assumed to reside exclusively on top of the grain mantle. It should 
hence be limited to $N_\text{act}$. We follow mechanism of \citet{Hincelin:2015} to account 
for desorption of any additional H$_2$. H$_2$ molecules that diffuse on top of an H$_2$ 
molecule in a layer below is assumed to desorb. Because of the fix we apply for surface 
reactions the additional term in the rate equation deviates from the original expression 
in \citet{Hincelin:2015}
\begin{equation}
 R_{\rm{H_2\, desorb}} = \exp\left(-\frac{E_{\rm{bind,H_2\, to\ ,H_2}}}{k_\text{B}T}\right) k_\text{diff} \frac{N_\text{act}^2}{N_\text{grain}} \theta \left(\frac{n_\text{s}(\rm{H_2})}{N_\text{act}}-\theta\right) 
\end{equation}
where
\begin{equation}
 \theta = 
\begin{cases}
    \frac{n_\text{s}(\rm{H_2})}{N_\text{act}},& \text{if } N_\text{H} <  N_\text{act}\\
    \frac{n_\text{s}(\rm{H_2})}{N_\text{H}},              &  \text{otherwise}
\end{cases}
\end{equation}
and  $N_\text{H} = n_\text{s}(\text{H}) + n_\text{s}(\rm{H_2})$. Choosing $E_{\rm{bind,H_2\, to\,H_2}}$, 
results in a maximum coverage of $N_\text{act}$ for H$_2$.

\end{document}